\documentclass[12pt,letterpaper]{article}
\usepackage[left=1in,top=1in,right=1in,nohead]{geometry}
\usepackage{enumerate}
\usepackage{graphicx}
\usepackage{amsmath}
\usepackage{amsthm}
\usepackage{mathrsfs}
\usepackage{amssymb}
\usepackage{multirow}
\usepackage{color}
\usepackage{bm}
\usepackage{comment}
\usepackage{subfigure}
\usepackage{paralist}
\usepackage{pstricks}
\usepackage{pst-coil,pstricks-add,pst-plot}
\usepackage{jheppub}
\usepackage{customcommands}

\newcommand{\Evac}{E_\mathrm{vac}}

\title{On Universality of Holographic Results for $(2+1)$-Dimensional CFTs on Curved Spacetimes}

\author{Sebastian Fischetti}
\author{and Toby Wiseman}

\affiliation{Theoretical Physics Group, Blackett Laboratory, Imperial College, London SW7 2AZ, UK}

\emailAdd{s.fischetti@imperial.ac.uk}
\emailAdd{t.wiseman@imperial.ac.uk}

\abstract{
The behavior of holographic CFTs is constrained by the existence of a bulk dual geometry.  For example, in~$(2+1)$-dimensional holographic CFTs living on a static spacetime with compact spatial slices, the vacuum energy must be nonpositive, certain averaged energy densities must be nonpositive, and the spectrum of scalar operators is bounded from below by the Ricci scalar of the CFT geometry.  Are these results special to holographic CFTs?  Here we show that for perturbations about 
appropriate backgrounds, they are in fact universal to \textit{all} CFTs, as they follow from the universal behavior of two- and three-point correlators of known operators.  In the case of vacuum energy, we extend away from the perturbative regime and make global statements about its negativity properties on the space of spatial geometries.  Finally, we comment on the implications for dynamics which are dissipative and driven by such a vacuum energy and we remark on similar results for the behavior of the Euclidean partition function on deformations of flat space or the round sphere.}

\begin{document}

\maketitle
\flushbottom

%================================== Introduction ==================================
\section{Introduction}
\label{sec:intro}

The concept of universality is a remarkably useful tool for gaining deeper insights into physical systems, as it identifies physical behaviors that are irrelevant of the details of any particular system.  For example, any relativistic field theory sufficiently close to a fixed point of the RG flow is expected to behave like a conformal field theory (CFT), and thus studying CFTs should give insights into the near-fixed-point behavior of a wider class of theories.  Indeed, a primary advantage of this simplification is that the conformal symmetry of CFTs strongly constrains their behavior, thereby giving rise to a large number of universal properties.  However, there may exist other universal features of CFTs which do not follow in any obvious way from conformal symmetry; it is these sorts of features that serve as the main topic of this paper.

How may we gain insight into such ``nontrivial'' universal features of CFTs?  One direction is provided by the AdS/CFT correspondence~\cite{Mal97,Wit98a,GubKle98}, which relates the behavior of certain CFTs to gravitational physics in asymptotically locally AdS spacetimes.  In particular, certain CFT states (in the limit of large central charge and strong coupling) are well-approximated by a classical gravitational dual, from which it is relatively easy to extract CFT statements.  The properties of interest are those that are robust in the sense that they apply to all, or at least to a wide class of holographic CFTs -- such as all those with duals that admit a conventional gravity description (in the sense of two derivative Einstein gravity coupled to matter).  In such a case, the CFT is strongly coupled, and thus these statements are typically very nontrivial from a field-theoretic perspective.  However, for this same reason we should not expect that all (or even most) of these statements should be universal to all (non-holographic) CFTs.  For instance, the celebrated universality of the shear viscosity to entropy density ratio,~$\eta/s = 1/4\pi$, only holds in CFTs with Einstein gravity duals~\cite{PolSon01,BucLiu03}; finite-coupling or finite-$N$ effects spoil this property, and in fact may even violate the conjectured bound~$\eta/s \geq 1/4\pi$ of~\cite{KovSon03,KovSon04} (see~\cite{Cre11} for a review).

The spirit of our approach is therefore the following: while we should not expect all properties of holographic CFTs with classical gravity duals to be universal, any sufficiently robust properties are candidates that may generalize beyond the holographic setting.  It is therefore natural to take such properties as guides for a purely field-theoretic analysis to see if they generalize to non-holographic CFTs.  There are a number of holographic results with the potential to generalize to universal statements (e.g.~the ``bound on chaos''~\cite{MalShe15} or~\cite{KoeLei15}'s holographic proof of the quantum null energy condition~\cite{BouFis15a,BouFis15b}, recently proven without holography in~\cite{BalFau17}) for CFTs on Minkowski spacetime.  While these are fascinating, they will not fall into the scope of this paper.  Instead, our focus will be on CFTs in \textit{curved} spacetimes.

Curved spacetime QFT is particularly challenging since it lacks the symmetry of the full Poincar\'e group present when working on flat space.  However, in the holographic context putting the CFT on a curved spacetime is very natural, changing only  the boundary conditions for the dual gravitational system. The dual problem remains a geometric one \cite{Marolf:2013ioa}, the tools to study it being essentially the same as for a holographic CFT on Minkowski spacetime. The result is that one can hope to have far greater control over a holographic CFT when the spacetime it resides on is curved than in the conventional CFT setting.  Robust statements for holographic CFTs can then be made that are surprising from purely CFT considerations.

Note that in this context, by a ``CFT'' we mean a theory on a general spacetime that is invariant under local conformal (or Weyl) transformations of the metric; such a theory should be more properly thought of as defined on a \textit{conformal} geometry.  In the special case where the conformal geometry is conformally Minkowski, we recover the usual notion of a CFT as a QFT invariant under the conformal group.  Since we will generally be interested in odd dimensions, there is no Weyl anomaly and we expect any CFT on flat space can be promoted to such a conformally (or Weyl) invariant theory on a general conformal geometry.

Let us now outline our results.  We will focus specifically on generalizing the curved space results of~\cite{HicWis15a,HicWis15b,FisHic16}, which found that if a holographic (2+1)-dimensional CFT on~$\rnum \times \Sigma$ (with~$\Sigma$ a compact space of genus no higher than one) has an Einstein gravity dual the following statements hold: (i) the vacuum energy~$\Evac$ (or Casimir energy) is non-positive; (ii) the vacuum energy density~$\rho_\mathrm{vac}$ is positive somewhere but negative when integrated inside any level set of the Ricci scalar of~$\Sigma$; and (iii) when~$\Sigma$ has spherical topology, the spectrum of a scalar operator~$\Phi$ is bounded below by the minimum value of the Ricci scalar of~$\Sigma$.  In fact, though these are derived for compact~$\Sigma$, results (i) and (ii) also hold on non-compact spaces, as can be inferred by taking~$\Sigma$ to be a torus whose size is taken arbitrarily large.  By performing a field theoretic analysis, we will show perturbatively that statements (i) and (ii) are indeed universal to all unitary (2+1)-dimensional CFTs when~$\Sigma$ is a small deformation of the round sphere or flat space\footnote{
% FOR ARXIV VERSION ONLY
In a previous version of this paper we considered a torus rather than flat space, incorrectly stating that a CFT on a periodically deformed flat space was equivalent to that same CFT on a torus.  This is obviously not the case; for it to be true the CFT fields would also have to be periodically identified.  We would thank Alexandre Belin and Henry Maxfield for pointing out this error.}, and that (iii) holds universally for the deformed round sphere.  Moreover, we will partially extend our perturbative result for property (i) to a global statement by relating the presence of geometries with positive~$\Evac$ to the behavior of~$\Evac$ at the boundary of the space~$\Mcal$ of geometries on which the theory is well-defined.  We emphasize again that while our analysis is motivated by holographic considerations, our final results are purely field theoretic and apply to all unitary (2+1)-dimensional CFTs.

It is worth highlighting explicitly the central theme of our perturbative approach.  The idea is to show that when~$\Sigma$ is a small deformation of the round sphere or flat space, the object of interest in properties (i)-(ii) above (namely, the vacuum energy density~$\rho_\mathrm{vac}$) is perturbed by a two-point correlator whose behavior is universal due to conformal symmetry.  Specifically, by working with the partition function~$Z$ in Euclidean signature, we show that the leading-order variation to the vacuum energy~$\Evac$ and energy density~$\rho_\mathrm{vac}$ are determined by the two-point function of the stress tensor on the unperturbed background geometry (i.e.~the round sphere or flat space).  Since for a unitary CFT this correlator is fixed (up to overall positive constant) by conformal invariance, then any CFT must reproduce the holographic results (i)-(ii) for small perturbations of $\Sigma$ away from the round sphere or flat space.  Likewise, when~$\Sigma$ is a small deformation of the round sphere the leading order perturbation to the lowest mode in the spectrum of any scalar operator~$\Phi$ can be related to a three point-correlator of~$\Phi$ with the stress tensor, which also has universal behavior. Thus the holographic result (iii) also holds for any CFT when~$\Sigma$ is a small perturbation of the sphere.  An important subtlety is to show that these results, which are all bounds and are saturated on the unperturbed space, are not saturated at this universal leading order in perturbation theory. If they were, higher orders which involve non-universal higher point functions could spoil the bounds.

We then extend these results slightly away from the perturbative regime and obtain a global statement on~$\Evac$.  Specifically, we will conclude that the neighborhood of the round sphere or flat space on which~$\Evac$ is negative must in fact extend to the boundary of the space of metrics on which the CFT is well-defined (that is, the space of conformal geometries with negative~$\Evac$ is ``large'').  This result makes it tantalizing to conjecture that for all $(2+1)$-dimensional unitary CFTs,~$\Evac$ may be non-positive on \textit{any} geometry~$\Sigma$.

In fact, these arguments make it natural to also consider the partition function~$Z$ on a compact Euclidean space that does not have a Lorentzian continuation. In particular, holographic arguments again bound $\ln Z$ for a three-dimensional holographic Euclidean CFT given a gravity dual with given topology \cite{And00,Mal11,HicWis15b}. For a round Euclidean sphere such a CFT saturates the bound, and any deformation to the sphere increases~$\ln Z$.  As above, we show that this result also extends perturbatively to arbitrary CFTs.  Such a statement is interesting in light of recent work on the so-called F-theorem, which conjectures that the partition function of Euclidean CFTs on a round sphere increases under RG flow \cite{JafKle11,Klebanov:2011gs}. Indeed the behavior of $\ln Z$ under particular squashing deformations of the sphere has been recently studied in \cite{Bobev:2016sap,BobBue17}. Another intriguing connection is that $\ln Z$ for a Euclidean $3$-dimensional CFT has been conjectured to describe the wavefunction of the universe \cite{Hertog:2011ky}, and the behavior of it on deformations of a round sphere then describes the probabilities associated to inhomogeneous cosmologies.

This paper is organized as follows.  In Section~\ref{sec:holog}, we review the aforementioned holographic bounds on the vacuum energy and spectrum of scalar operators, which follow purely from bulk geometric statements.  In Section~\ref{sec:PT} we use perturbation theory to show that the bounds on the vacuum energy (and energy density) are obeyed perturbatively in any (2+1)-dimensional CFT under static perturbations to flat space or the round sphere.  In Section~\ref{sec:nonpert} we extend the perturbative bound on~$\Evac$ into global statements and study some of the physical consequences of this result.  Finally, in Section~\ref{sec:spectrum} we show that the bound on the spectrum of scalar operators is also obeyed perturbatively in any CFT under static perturbations to flat space or the round sphere, and we briefly summarize in Section~\ref{sec:summary}.  Appendices~\ref{app:flatpert} and~\ref{app:spherepert} generalize our perturbative results for~$\Evac$ to analogous perturbative bounds on $\ln Z$ for deformations of Euclidean flat space or the three-sphere.

\textbf{Notations and Conventions.}  Unless otherwise specified, we will consider CFTs on a three-dimensional Euclidean geometry~$M$ which we take to be of the form~$\rnum \times \Sigma$ for some two-dimensional space~$\Sigma$.  The Euclidean metric on~$M$ will be denoted by~$G_{AB}$ and the metric on~$\Sigma$ by~$g_{ij}$.  We will sometimes invoke a holographic bulk dual; spatial indices (i.e.~non-Euclidean time indices) in this dual will be denoted by~$I,J,\ldots$.  We warn the reader that some of these conventions differ from those of~\cite{HicWis15a,HicWis15b,FisHic16}.

%============================== Holographic arguments  ============================
\section{Review of Holographic Bounds}
\label{sec:holog}

Let us now briefly review the holographic derivation of the bounds advertised above; more details can be found in~\cite{HicWis15a,HicWis15b,FisHic16}.

\subsection{$\Evac \leq 0$}
\label{subsec:holoEbound}

First, we review the negativity of the vacuum energy of a holographic CFT on~$\rnum \times \Sigma$ with a gravity dual when $\Sigma$ is taken to be compact. 
Because the boundary metric is static, we expect that the bulk dual to the vacuum state should be static as well.  It is therefore convenient to work in Euclidean signature and consider a general three-dimensional Euclidean boundary metric~$G_{AB}$.  Recall that per the usual holographic dictionary, a classical bulk solution corresponds to a saddle point of the Euclidean partition function~$Z$, and therefore we have that in the leading semiclassical approximation (or $1/N$ expansion when the CFT is a gauge theory, such as the ABJM theory \cite{Aharony:2008ug}) the on-shell Euclidean action~$S_E$ of the bulk determines the partition function as~$\ln Z = -S_E$.

We will work in the universal gravity sector, where the bulk is described by vacuum Einstein gravity.  Note that even if the bulk dual to the CFT vacuum is not contained in this sector (e.g.~if the bulk dual to the vacuum state contains condensates of bulk matter fields), any state described by pure gravity will have an energy no greater than that of the true vacuum~$\Evac$.  Thus showing that~$\Evac \leq 0$ in the universal gravity sector is sufficient to show that~$\Evac \leq 0$ for all vacuum states on geometries admitting pure gravity bulk duals.

Now, in this universal sector, the (suitably regulated) on-shell action is proportional to the renormalized volume~$V_\mathrm{ren}$ of the bulk:~$V_\mathrm{ren} = 8\pi G_N l^2 S_E /3$, with~$G_N$ the bulk Newton's constant and~$l$ the AdS scale~\cite{Mal11}.  The partition function is therefore related to this renormalized volume as
\be
\ln Z = -\frac{3}{8\pi G_N l^2} V_\mathrm{ren}.
\ee
The key observation stems from a result of Anderson~\cite{And00}: the Gauss-Bonnet theorem implies that the renormalized volume of the bulk is bounded by its topology as
\be
\label{eq:andersonbound}
V_\mathrm{ren} \leq \frac{4 \pi^2}{3} l^4 \chi_\mathrm{bulk},
\ee
with~$\chi_\mathrm{bulk}$ the bulk Euler characteristic.  The partition function therefore obeys the bound
\be
\label{eq:Zholobound}
\ln Z \geq -8\pi^2 c_\mathrm{holo} \chi_\mathrm{bulk},
\ee
where~$c_\mathrm{holo} \equiv l^2/(16\pi G_N)$ is the effective CFT central charge of the holographic theory (with~$c_\mathrm{holo} \sim N^2$ in the case of a gauge theory). We require~$c_\mathrm{holo} \gg 1$ in order to justify the semiclassical approximation of $\ln Z$ by the classical bulk dual.  It is worth noting that~\eqref{eq:andersonbound} (and therefore also~\eqref{eq:Zholobound}) is saturated if and only if the bulk Weyl tensor vanishes, implying that the bulk must (locally) be hyperbolic space.  

To prove that~$\Evac \leq 0$ for holographic CFTs on~$\rnum \times \Sigma$, it is convenient to first work with thermal states with 
inverse temperature~$\beta \equiv 1/T$ and then take the zero-temperature limit (subject to a mild regularity assumption).  We therefore temporarily place the CFT on~$S^1 \times \Sigma$, writing the boundary metric as\footnote{Fermions in the CFT are given antiperiodic boundary conditions around the~$S^1$.}
\be
ds^2_B = d\tau^2 + g_{ij}(x) dx^i \, dx^j
\ee
with~$\tau \sim \tau + \beta$ and~$g_{ij}$ the metric on~$\Sigma$.  We also assume that the~$U(1)$ isometry generated by~$\partial/\partial \tau$ is shared by the full bulk solution.  Now we make two observations.  First, the partition function determines the free energy~$F$ of the state as~$F = -T\ln Z$.  Second, the bulk Euler characteristic~$\chi_\mathrm{bulk}$ can be expressed in terms of the Euler characteristics of the fixed points (``bolts'') of the~$U(1)$ isometry (which correspond to Killing horizons in Lorentzian signature) as~\cite{Gibbons:1979xm,HicWis15b}
\be
\chi_\mathrm{bulk} = \sum_n \chi_n,
\ee
where~$n$ indexes the bolts and~$\chi_n$ are their Euler characteristics.  Combining these two observations with~\eqref{eq:Zholobound}, we obtain
\be
F \leq 8\pi^2 c_\mathrm{holo} T \sum_n \chi_n.
\ee
The desired bound on the vacuum energy~$\Evac$ is obtained by taking the~$T \to 0$ limit: noting that~$F = E - TS$ and assuming the ground state is sufficiently well-behaved that~$TS \to 0$ in this limit, we find that~$\Evac = \lim_{T \to 0} F$ and therefore
\be
\label{eq:Evacbound}
\Evac \leq 0,
\ee
with saturation only when the bulk is Euclidean AdS (or a quotient thereof).  It is worth emphasizing explicitly that since this bound is obtained from a classical bulk geometry, it is really a statement about the leading-order behavior of $\Evac \sim N^2$.  But since it is saturated only when the bulk is (locally) pure Euclidean AdS, it must in fact hold under arbitrary perturbative corrections in~$1/N$.

As an aside, it is interesting to note that the bound~\eqref{eq:Zholobound} holds for any Euclidean boundary geometry.  If we are uninterested in a static Lorentzian continuation, we may for instance take the boundary metric to be a topological three-sphere.  We would then expect the bulk vacuum geometry to have Euler characteristic~$\chi_\mathrm{bulk} \leq 1$, with~$\chi_\mathrm{bulk} = 1$ in the most natural case where the bulk is a topological four-ball (though note it is unclear that a bulk infilling solution should always exist; see~\cite{Anderson:2001pf} for work in this direction).  We therefore conclude that for a holographic Euclidean CFT on a topological three-sphere, we have
\be
\label{eq:lnZBallbound}
\ln Z \geq -8\pi^2 c_\mathrm{holo}.
\ee
As for the energy, this bound is saturated only when the bulk is (locally) hyperbolic space, and so the boundary metric is the round three-sphere.

\subsection{Vacuum Energy Density}
\label{subsec:holorhobound}

Since~$\Evac$ is generically negative, it follows that the \textit{local} energy density~$\rho_\mathrm{vac}$ on~$\rnum \times \Sigma$ must be negative somewhere as well.  Can this statement be made more precise?

The answer is yes \cite{FisHic16}, though the simple topological arguments used above are insufficient for doing so.  The idea is to decompose the bulk metric into the so-called optical metric~$\bar{g}_{IJ}$ on each (Euclidean) time slice as
\be
\label{eq:bulkoptical}
ds^2_\mathrm{bulk} = \frac{l^2}{Z^2(y)} \left(d\tau^2 + \bar{g}_{IJ}(y) dy^I dy^J \right),
\ee
where~$Z(y)$ is essentially a redshift factor.  The equations of motion and Bianchi identities then imply that the Ricci scalar computed from the optical metric~$\bar{g}_{IJ}$ obeys
\be
\label{eq:opticalR}
\overline{\grad}^2 \bar{R} = -3\left(\bar{R}_{IJ} - \frac{1}{3} \bar{R} \, \bar{g}_{IJ}\right)^2 \leq 0,
\ee
from which it follows that for an arbitrary non-negative and non-increasing function~$\bar{f}$, we must have
\be
\label{eq:Rfoptical}
\overline{\grad}^I \left(\bar{f}(\bar{R}) \overline{\grad}_I \bar{R} \right) = \bar{f}(\bar{R}) \overline{\grad}^2 \bar{R} + \bar{f}'(\bar{R}) \left(\overline{\grad} \bar{R}\right)^2 \leq 0.
\ee
Integrating the left-hand side of~\eqref{eq:Rfoptical} over an entire bulk time slice, integrating by parts, using the near boundary behavior of~$\bar{R}$, and imposing appropriate well-behavedness properties of the bulk in the~$T \to 0$ limit, one obtains~\cite{FisHic16}
\be
\label{eq:frhobound}
\int_\Sigma f(R) \rho_\mathrm{vac} \leq 0,
\ee
where~$f$ is any non-negative and non-increasing function (simply related to~$\bar{f}$ by~$f(x) = \bar{f}(3x)$),~$R$ is the Ricci scalar of~$\Sigma$, and the usual volume form on $\Sigma$ is implied.  In the limiting case that~$f(x)$ is taken to be a step function, it follows that the vacuum energy density obeys the family of bounds
\be
\label{eq:rhobound}
\int_{R \leq r} \rho_\mathrm{vac} \leq 0 \quad \forall r.
\ee
Note that taking~$r \geq \max(R)$ then simply reproduces the bound~\eqref{eq:Evacbound}, while taking~$r \to \min(R)$ implies that~$\rho_\mathrm{vac} \leq 0$ at a point where~$R$ attains its minimum value.  Again we emphasize that since they were derived from a classical bulk, all these statements are really constraints on the~$O(N^2)$ behavior of~$\rho_\mathrm{vac}$ in~$N$.  However, since all of these inequalities are only saturated when the bulk is pure AdS (or a quotient theoreof), for nontrivial boundary geometry they are strict.  This implies that they are in fact robust to perturbative corrections in~$1/N$.  However, unlike the bound $\Evac \le 0$, these local energy density bounds are less robust if~$\Sigma$ is sufficiently deformed that the true vacuum involves condensates of other bulk fields (i.e.~if the bulk dual to the vacuum is not in the universal gravity sector).

While the bounds~\eqref{eq:rhobound} are formally derived for compact~$\Sigma$, they hold just as well when~$\Sigma$ is a deformation of flat space.  This can be intuited by taking~$\Sigma$ to be a flat torus deformed in some region~$\mathcal{R}$.  Taking the size of the torus to be arbitrarily large while keeping~$\mathcal{R}$ fixed, we expect~$\rho_\mathrm{vac}$ to become independent of the torus size in this limit and to vanish outside of~$\mathcal{R}$.  Thus we conclude that~\eqref{eq:rhobound} (as well as~\eqref{eq:Evacbound}) is obeyed when~$\Sigma$ is any compactly-supported deformation of flat space.  More precisely,~\eqref{eq:rhobound} can be rederived for non-compact~$\Sigma$ using essentially the same bulk gravity derivation as in the compact case with only two key differences.  First, the deformation of~$\Sigma$ from flat space must fall off sufficiently fast that~$\int f(R) \rho_\mathrm{vac}$ is finite.  Second, we must work directly at zero temperature (rather than take a zero-temperature limit) to ensure that any potential boundary terms from bulk horizons vanish; this will be guaranteed assuming the bulk ends on an extremal horizon with finite entropy\footnote{The virtue of working with the zero-temperature limit for compact~$\Sigma$ is that it allows the IR to have a ``good'' singularity \cite{Gubser:2000nd}.  However, in the non-compact setting, at finite temperature the boundary term from the horizon will typically diverge.  To avoid this, one is restricted to only consider zero temperature from the outset.  In this sense for non-compact $\Sigma$ the bounds are somewhat less robust as they do not cover the possibility that the bulk ends in a good IR singularity, although we expect that this does not occur and the IR is a regular extremal horizon.}.  As in the compact case, the bounds will be sharp only for flat space -- any localized  deformation of it will result in strict inequalities.

Finally, we mention that despite the fact that the family of bounds~\eqref{eq:rhobound} implies \textit{negativity} conditions on~$\rho_\mathrm{vac}$, it is in fact possible to show that when~$\Sigma$ has genus zero or one,~$\rho_\mathrm{vac}$ must generically be \textit{positive} somewhere (with the only exception again being pure AdS, when~$\rho_\mathrm{vac} = 0$ everywhere).  The proof of this statement follows from the so-called inverse mean curvature flow \cite{FisHic16} and is essentially a consequence of the results of~\cite{LeeNev15}.

\subsection{Spectrum of Scalar Operators}
\label{subsec:holoscalars}

Let us conclude by reviewing a final bound on the spectrum of scalar operators.  Specifically, consider a CFT of any dimension~$d$ on a static spacetime~$\rnum \times \Sigma$ for compact~$\Sigma$, and consider a scalar operator~$\Ocal$ of dimension~$\Delta$\footnote{We note that the scaling dimension $\Delta$ of a CFT operator in Minkowski spacetime is defined from its transformation under dilatations. For a Weyl invariant theory on a curved background, the scaling dimension of a local operator is similarly defined from its behavior under Weyl transformation.} dual to a minimally coupled bulk scalar field~$\phi$ of mass~$l^2 m^2 = \Delta (d - \Delta)$.  Assuming the CFT vacuum state corresponds to a static bulk spacetime with no horizons or singularities, the bulk field~$\phi$ may be expanded in the optical coordinates of~\eqref{eq:bulkoptical} as~$\phi = e^{i \omega t} Z^\Delta j_\omega$ for some~$j_\omega(y^I)$.  In~\cite{HicWis15a} it was shown that the bulk equation of motion~$\grad^2_\mathrm{bulk}\phi = m^2 \phi$ can be used to express the frequency~$\omega$ in terms of~$j_\omega(y^I)$ as
\be
\label{eq:omegaexpr}
\omega^2 = \left[\int \sqrt{\bar{h}} \, Z^{2 \Delta + 1 - d} j_\omega^2\right]^{-1}\int  \sqrt{\bar{h}} \, Z^{2 \Delta + 1 - d} \left( \left( \partial j_\omega \right)^2 + \frac{\Delta^2}{d(d-1)} \bar{R} j_\omega^2 \right),
\ee
where the integrals are taken over a static time slice of the bulk.  We emphasize that this expression is written covariantly with respect to the optical metric $\bar{g}_{IJ}$ (in particular,~$\bar{R}$ is the Ricci scalar of~$\bar{g}_{IJ}$).

In~\cite{HicWis15a}, the above expression for~$\omega^2$ was used to obtain a lower bound as follows.  In general dimension, the generalization of the elliptic equation~\eqref{eq:opticalR} can be used to conclude that~$\bar{R}$ is minimized on the asymptotic boundary~$Z = 0$, where an analysis of the asymptotics shows that it is related to the boundary Ricci scalar~$R$ as~$\bar{R} = dR/(d-2)$.  This fact, coupled with positivity of the first term in~\eqref{eq:omegaexpr} above, guarantees that the lowest frequency in the spectrum is bounded below by 
\be
\label{eq:specbound}
\omega^2_\mathrm{min} \ge \frac{\Delta^2}{(d-1)(d-2)} R_\mathrm{min},
\ee
with~$R_\mathrm{min}$ the minimum value of the boundary Ricci scalar (i.e.~the Ricci scalar of~$\Sigma$).  Note that this bound is saturated precisely when~$\Sigma$ is a round sphere, in which case the lowest mode has constant~$j_\omega$.

Let us pause for a few comments.  First, the reader may be concerned that this bound is not ``universal'' in the sense that even in the purely holographic setting, it does not arise in  the pure gravity sector of AdS/CFT: it requires the addition of the bulk scalar field~$\phi$.  But generically any top-down construction of a holographic duality will involve many such bulk scalars (these might come from e.g.~supergravity scalars or from Kaluza-Klein modes of an internal space); while such scalars may have complicated nonlinear actions, for fluctuations about the ``universal vacuum'' where only gravity is turned on these fluctuations obey the canonical linear scalar field equation~$\grad^2_\mathrm{bulk} \phi = m^2 \phi$.  The bound above would then apply to such fluctuations.

Secondly, we emphasize that the holographic calculation requires~$\Sigma$ to be compact and that the bulk dual to the vacuum state have no horizons or singularities.  But when~$\Sigma$ has genus higher than a sphere, the bulk vacuum geometry generically ends on a (quotient of an) extremal horizon.  We therefore expect the bound to only apply when~$\Sigma$ has the topology of a sphere; in such a case, the bound can be interpreted as a guarantee that the vacuum geometry is stable against scalar fluctuations, and hence against scalar condensation, as long as the Ricci curvature remains positive everywhere.

Thirdly, the bound~\eqref{eq:specbound} is not just an artefact of our holographic derivation; indeed, we can obtain the same bound for a free conformal scalar.  To that end, consider a free conformal scalar operator~$\Phi$ living on~$\rnum \times \Sigma$ (still with $\Sigma$ compact); this operator obeys
\be
\label{eq:freeconformal}
\left(-\partial_t^2 + \grad^2 \right) \Phi = \frac{d-2}{4(d-1)} R \Phi
\ee
where~$\grad^2$ and~$R$ are the Laplacian and scalar curvature of~$\Sigma$.  We may again expand~$\Phi$ into modes as
\be
\Phi = e^{i \omega t} J_\omega
\ee
for some wave function~$J_\omega$.  With this decomposition,~\eqref{eq:freeconformal} takes the form
\be
\label{eq:freeconformalspectrum}
\omega^2 J_\omega = \left(-\grad^2 + \frac{d-2}{4(d-1)} R \right) J_\omega.
\ee
Multiplying by~$J_\omega$, integrating over~$\Sigma$, and integrating by parts using the fact that~$\Sigma$ is compact, we obtain
\be
\label{eq:freeconformalomega}
\omega^2 = I[J_\omega] \equiv \left[\int_\Sigma J_\omega^2\right]^{-1} \int_\Sigma \left( (\partial J_\omega)^2 + \frac{d-2}{4(d-1)} R J_\omega^2\right)
\ee
(the natural volume element on~$\Sigma$ is understood).  Again noting positivity of the first term in the integral and the fact that~$\Phi$ has conformal weight~$\Delta = d/2-1$, we thus immediately recover the bound~\eqref{eq:specbound}.  That the bound is obeyed for a holographic and a free scalar CFT is very suggestive that it is likely to hold more generally, and indeed we shall later provide evidence that this is the case in any CFT for perturbations of the round sphere.

Finally it is interesting that the bound~\eqref{eq:specbound} is given in terms of the Ricci scalar, when perhaps the na\"ive expectation for such a bound would be in terms of the inverse volume of $\Sigma$. Could there be a ``simpler'' bound of the form $\omega_\mathrm{min}^2 \ge c \mathrm{Vol}(\Sigma)$ for a positive constant $c$ universal to all CFTs?  In fact in the case of the free scalar it is easy to show that for smooth (and hence finite volume) $\Sigma$ it is possible for~$\omega_\mathrm{min}^2$ to be negative (provided that $R_\mathrm{min}$ is), ruling out such a bound. Consider a simple example taking~$\Sigma$ to be a deformed sphere:
\be
\label{eq:deformedsphere}
ds^2_\Sigma = d\theta^2 + \sin^2 \theta \left( 1 + r \sin^2 \theta \right)^2 d\phi^2,
\ee
where~$r > -1$ to ensure the perturbation is regular everywhere.  The Ricci scalar is
\be
R =  \frac{2 - 3 r \left( 1 + 3 \cos{2\theta}\right) }{1 + r \sin^2{\theta}},
\ee
and thus for~$r > 1/6$ we see $R < 0$ near the poles~$\theta = 0$,~$\pi$.  Now, by~\eqref{eq:freeconformalomega} and standard variational arguments, any test wave function~$J_\mathrm{test}$ yields a bound~$\omega^2_\mathrm{min} \leq I[J_\mathrm{test}]$\footnote{As a reminder, the idea is to note that any trial wave function~$J_\mathrm{test}$ can be decomposed into a linear combination of the orthonormal eigenmodes~$J_\omega$.  Then~$I[J_\mathrm{test}]$ reduces to a weighted average over the eigenfrequences~$\omega^2$, implying that~$I[J_\mathrm{test}]$ cannot be less than the lowest such frequency.}.  Specifically, taking
\be
J_\mathrm{test}(\theta,\phi) = \frac{1 + \cos^2 \theta}{2}
\ee
we find
\be
\omega^2_\mathrm{min} \leq I[J_\mathrm{test}] = \frac{7(15-4r)}{4(49+26r)},
\ee
and thus for~$r > 15/4$ we are guaranteed that~$\omega^2_\mathrm{min} < 0$.  This verifies that any lower bound on~$\omega^2_\mathrm{min}$ must be allowed to go negative, and in particular cannot be as simple as an inverse volume scaling.  We will return to this example later in section~\ref{subsec:lnZproperties} as it shows this scalar CFT partition function may become ill-defined for smooth $\Sigma$.

%================================= Perturbation theory  =================================
\section{Perturbations of (Locally) Weyl Flat Metrics}
\label{sec:PT}

Let us now study how the holographic results outlined above are reproduced perturbatively in general CFTs.

\subsection{$\Evac$ and the Partition Function}
\label{subsec:partfuncproperties}

We are ultimately interested in Lorentzian CFTs on static spacetimes, but as above it is convenient to work in Euclidean signature.  In that case, recall that for a general QFT, the partition function is defined on a given Euclidean space~$M$ with metric~$G_{AB}$ as~$Z[G] = \int \Dcal X e^{-S_E[G] }$, where~$S_E[G]$ is the Euclidean action and~$X$ are all the QFT fields being integrated over in the path integral.  This partition function defines the vacuum state in the sense that its functional derivatives yield the vacuum expectation values (VEVs) of all QFT operators.  In particular, the VEV of the stress tensor operator~$T_{AB} \equiv (2/\sqrt{G}) \delta S_E/\delta G^{AB}$ is given by
\be
\ev{T_{AB}}_G = -\frac{2}{\sqrt{G}} \frac{\delta \ln Z[G]}{\delta G^{AB}},
\ee
and so the variation to~$\ln Z$ under a perturbation~$G_{AB} \to G_{AB} + \eps H_{AB}$ is given by
\be
\label{eq:linvariation}
\ln Z[G + \eps H] = \ln Z[G] + \frac{\eps}{2} \int_M  \ev{T_{AB}}_G H^{AB} + O(\eps^2),
\ee
where indices are raised and lowered with respect to the background metric~$G_{AB}$ and here and below, the natural volume element on~$M$ is understood in the integral.

Here we are specifically interested in a CFT living on Euclidean~$\rnum \times \Sigma$ with~$\Sigma$ a two-dimensional space.  However, for the purposes of this section it will again be convenient to instead take the CFT to live on the Euclidean geometry~$M = S^1 \times \Sigma$, whose metric we write as
\be
\label{eq:metric}
ds^2 \equiv G_{AB} dx^A dx^B = d\tau^2 + g_{ij}(x) dx^i dx^j
\ee
with~$g_{ij}$ the metric on~$\Sigma$ and the Euclidean time~$\tau$ periodic with period~$\beta$.  With appropriate fermion boundary conditions we may regard this continuation as the usual one for thermal field theory, with~$\beta$ being inverse temperature.  Then (under the assumption that the system is well-behaved at zero temperature and has a non-degenerate vacuum state) our vacuum results can be obtained in the~$\beta \to \infty$ limit, which sends the~$S^1$ to~$\rnum$.  Specifically, we are interested in the vacuum energy~$\Evac$, defined as
\be
\label{eq:Evacdef}
E_\mathrm{vac} = - \lim_{\beta \to \infty} \frac{\partial}{\partial \beta} \ln Z.
\ee
In this limit we expect the ground state to dominate the dynamics, and therefore the free energy~$F = -\beta^{-1} \ln Z$ should approach~$\Evac$:~$F \to \Evac$.  This observation is crucial, as it allows us to exploit features of the partition function to derive properties of~$\Evac$.  Let us therefore compile these properties before deriving our perturbative results.

From the Lorentzian perspective the metric~$G_{AB}$ is static, and therefore we expect the VEV of the stress tensor to be static as well. 
This implies its only nonzero components are the vacuum energy density~$\rho$ and spatial stress tensor~$t_{ij}$,
\be
\rho[G] = - \ev{T_{\tau\tau}}_G, \quad  t_{ij}[G] = \ev{T_{ij}}_G,
\ee
and that these are both constant in~$\tau$.  In particular, note that a constant perturbation~$\delta G^{\tau\tau}$ changes~$\beta$ by~$\delta \beta = -\delta G^{\tau\tau} \beta/2$, and thus per definition~\eqref{eq:Evacdef} we may write~$\Evac$ as the integral of~$\rho$ on~$\Sigma$:
\be
\label{eq:Evacrho}
\Evac =  -\lim_{\beta \to \infty} \int d\tau \int_{\Sigma} \frac{1}{\sqrt{g}} \frac{\delta \ln Z}{\delta G^{\tau\tau}} \frac{\delta G^{\tau\tau}}{\delta \beta} = \int_\Sigma \rho,
\ee
where we used the fact that~$\int d\tau = \beta$.  Likewise, we may also consider a purely spatial perturbation~$h_{ij}$ which leaves~$G_{\tau\tau}$ unchanged; then from~\eqref{eq:linvariation} we have
\be
\ln Z[g+ \eps h] = \ln Z[g] + \frac{\eps}{2} \, \beta \int_\Sigma t_{ij}[g] h^{ij} + O(\eps^2),
\ee
and therefore again from~\eqref{eq:Evacdef} we find that
\be
\label{eq:secondvarElin}
\Evac[g + \epsilon h] = \Evac[g] - \frac{\eps}{2} \int_\Sigma t_{ij}[g] h^{ij} + O(\eps^2).
\ee
We may interpret this expression as a generalization of the usual thermodynamic relation ``$\delta E = - P \delta V$''.  Moreover, having now exploited the~$\beta \to \infty$ limit to obtain~\eqref{eq:Evacrho} and~\eqref{eq:secondvarElin}, we no longer have a use for the regulator~$\beta$, and henceforth we will proceed at zero temperature.

Our goal will be to derive universal bounds on~$\Evac[g + \eps h]$ when~$g$ is  flat space or the round two-sphere.  In order to do so, we must express~$\ev{T_{AB}}_{G + \eps H}$ in terms of universal quantities.  Let us thus restrict to~$(2+1)$-dimensional CFTs and consider the behavior of~$\ln Z$ under perturbations when~$G_{AB} = \omega^2(x) \delta_{AB}$ is (locally) conformally flat.  In particular, we will only consider theories on~$G_{AB}$ whose ground states are obtained via the conformal map generated by~$\omega$ from the ground state on Euclidean flat space~$\rnum^3$; in other words, all VEVs in the ground state on~$M$ must be obtained via conformal transformations from the VEVs of the ground state on flat space\footnote{This requirement is important: for instance, if the Weyl factor~$\omega(x)$ is singular or vanishes somewhere, the geometry~$G_{AB}$ fails to be \textit{globally} conformally flat.  In such a case, one could in principle consider a CFT on~$G_{AB}$ at which ``extra'' boundary conditions are imposed at these singularities; the ground state of such a theory would not be obtained in any simple way from the ground state on~$\rnum^3$, and the universality results discussed here would not obviously apply.}.  In such a theory, the ground state must be invariant under the conformal group, which in particular means that one-point functions must vanish.  The expansion of the partition function thus yields
\begin{multline}
\label{eq:lnZvariation}
\ln Z[G + \epsilon H] = \ln Z[G] \\ + \frac{\eps^2}{8} \int d^3x \sqrt{G(x)} \int d^3y \sqrt{G(y)} \, \ev{T_{AB}(x) T_{CD}(y)}_\omega H^{AB}(x) H^{CD}(y) + O(\eps^3),
\end{multline}
where the subscript~$\omega$ denotes that the expectation value is computed in the geometry~$G_{AB} = \omega^2(x) \delta_{AB}$\footnote{A discussion of this was given recently in \cite{BobBue17}, where the order $O(\eps^3)$ terms involving 3-point functions are given explicitly.}.  But the two-point function above is obtained via a conformal transformation from that on flat space, which is in turn heavily constrained by conformal symmetry.  Indeed, in the conformally flat coordinates in which the metric takes the form~$G_{AB} = \omega^2(x) \delta_{AB}$, we have
\begin{subequations}
\be
\label{eq:TTuniversal}
\ev{T_{AB}(x_1) T_{CD}(x_2)}_\omega = \omega^{-1}(x_1) \omega^{-1}(x_2) \frac{c_T}{| x_1 - x_2 |^6} I_{AB,CD}(x_1 - x_2),
\ee
with
\bea
I_{AB,CD}(x) &\equiv \frac{1}{2} \left( I_{AC}(x) I_{BD}(x) + I_{AD}(x) I_{BC}(x) \right) - \frac{1}{3} \delta_{AB} \delta_{CD}, \\
I_{AB}(x) &\equiv \delta_{AB} - \frac{2 x_A x_B}{ x^2 },
\eea
\end{subequations}
and where the overall normalization $c_T$ is required to be positive by unitarity. In the case that the CFT has a holographic dual,~$c_T$ is related to the ``holographic'' central charge as
\be
\label{eq:cTgrav}
\frac{\pi^2 c_T}{48} = c_\mathrm{holo} \equiv \frac{l^2}{16 \pi G_N} \, .
\ee

Next, in order to obtain a useful explicit expression for~$\Evac$, it is convenient to re-express~$\ln Z[G + \eps H]$ in terms of one-point functions which take a universal form.  To do so, note that by taking a functional derivative of~\eqref{eq:lnZvariation}, we conclude that the one-point function of the stress tensor in the \textit{perturbed} geometry is
\be
\label{eq:pertonepoint}
\ev{T_{AB}(x)}_{G + \eps H} = \frac{\eps}{2} \int d^3y \sqrt{G(y)} \ev{T_{AB}(x) T_{CD}(y)}_\omega {H}^{CD}(y) + O(\eps^2).
\ee
Therefore, for a conformally flat geometry~$G_{AB}$, the variation of~$\ln Z$ can be written as
\be
\label{eq:linvariation2}
\ln Z[G + \eps H] = \ln Z[G] + \frac{\eps}{4} \int d^3x  \sqrt{G(x)} \ev{T_{AB}(x)}_{G + \eps H} H^{AB}(x)  + O(\eps^3).
\ee
(The attentive reader may have noticed that the coefficient of the integral in this expression differs by a factor of~$-1/2$ from what might have been expected from~\eqref{eq:linvariation}.  This is due to the fact that the variation~$\eps H_{AB}$ and the geometry~$G_{AB} + \eps H_{AB}$ are not independent, and thus~\eqref{eq:linvariation} does not apply.)

Now, the unperturbed spacetimes in which we are interested take the form~$\rnum \times \Sigma$ for~$\Sigma$ either flat space or the round sphere.  Both of these are (locally) conformally flat, and therefore the change in the partition function under a perturbation of~$\Sigma$ is captured by~\eqref{eq:linvariation2}.  Considering therefore static perturbations~$\eps h_{ij}$ to the metric~$g_{ij}$ of~$\Sigma$ and using~\eqref{eq:linvariation2} and~\eqref{eq:Evacdef}, we finally obtain our expression for the perturbed vacuum energy:
\be
\label{eq:secondvarE}
\Evac[g + \eps h] = \Evac[g] - \frac{\eps}{4} \int_\Sigma t_{ij}[g + \eps h] h^{ij} + O(\eps^3),
\ee
where from~\eqref{eq:pertonepoint} we have
\begin{subequations}
\label{eq:lintijrho}
\bea
t_{ij}[g + \eps h](x) &= \frac{\eps}{2} \int d\tau \int d^2y \sqrt{g(y)} \, \ev{T_{ij}(0,x) T_{kl}(\tau,y)}_\omega h^{kl}(y) + O(\eps^2), \label{eq:lintij} \\
\rho[g + \eps h](x) &= -g^{ij} t_{ij}[g + \eps h](x) \label{eq:linrho}.
\eea
\end{subequations}
It is worth mentioning explicitly that while both~$t_{ij}$ and~$\rho$ are~$\tau$-independent, the two-point function of the stress tensor actively does depend on the difference in Euclidean time of its arguments.

\subsection{Perturbative Energy Bounds}
\label{subsec:pertbounds}

We may now derive the advertised bounds~\eqref{eq:Evacbound} and~\eqref{eq:rhobound} on~$\Evac$ and~$\rho_\mathrm{vac}$ for any $(2+1)$-dimensional unitary CFT for~$\Sigma$ a deformation of flat space or the round sphere.  These bounds are clearly saturated when~$\Sigma$ is flat or a round sphere (since then~$\rho_\mathrm{vac} = 0$), and from~\eqref{eq:secondvarE} and~\eqref{eq:lintijrho} it follows that the leading-order corrections to~$\Evac[g+\eps h]$ and~$\rho_\mathrm{vac}[g + \eps h]$ can be expressed in terms of the two-point function~\eqref{eq:TTuniversal} of the stress tensor.  Because this two-point function is universal to all CFTs, we may in particular compute this leading-order correction in holographic CFTs.  Showing that~\eqref{eq:Evacbound} and~\eqref{eq:rhobound} are obeyed perturbatively by all CFTs thus amounts to showing that they are obeyed as \emph{strict} inequalities at leading nontrivial order in~$\eps$ in holographic CFTs for \emph{any} perturbation $h_{ij}$ that deforms~$\Sigma$ from being flat or a round sphere.  One is then guaranteed that in some neighbourhood of geometries of flat space or the round sphere, the bounds are obeyed by any CFT\footnote{\label{footneighborhood}When referring to the ``neighborhood'' of a sphere, we will mostly be concerned with deformations that take one away from the \textit{round} sphere.  That is, we are uninterested in perturbations that induce rigid rescalings (since these do not change the conformal class of the full Euclidean geometry~$M$), as such perturbations leave the bounds saturated exactly nonperturbatively.  We will discuss this subtlety further when it becomes relevant.}.  We emphasize the importance of showing the bounds are strict for any perturbation: if there existed perturbations that left the bounds merely saturated at leading order in~$\eps$, then non-universal higher-point functions might spoil them at higher order in~$\eps$.

To show that the bounds are indeed strictly obeyed at leading order in~$\eps$, we rederive the holographic bound~\eqref{eq:frhobound} in the perturbative regime.  Let us first take~$f(x) = 1$, in which case by integrating over a static time slice~$\overline{\Sigma}$ of the optical bulk metric we have
\bea
\label{eq:EvacRsquared}
\Evac[g + \eps h] = \int_\Sigma \rho_\mathrm{vac} = \int_{\overline{\Sigma}} \overline{\grad}^2 \bar{R} = -3 \int_{\overline{\Sigma}} \left(\bar{R}_{IJ} - \frac{1}{3} \bar{R} \bar{g}_{IJ}\right)^2,
\eea
where we have used the fact that at zero temperature any contributions from horizons vanish, the last equality is obtained using~\eqref{eq:opticalR}, and the second is derived in~\cite{FisHic16}.  For $\Sigma$ a round sphere or flat space, the Ricci scalar $R$ of $\Sigma$ is constant and the bulk vacuum geometry is locally AdS; consequently the optical metric~$\bar{g}$ is Einstein,~$\Evac[g] = 0$ and the bound~\eqref{eq:frhobound} is saturated.  This is consistent with our previous discussion, where we know the vacuum energy density (being a one-point function) must vanish exactly, not just to leading order in large $c_\mathrm{holo}$.  For any perturbation $h$ to the metric $g$ on $\Sigma$ away from a round sphere or flat space, the Ricci scalar of $\Sigma$ will be perturbed away from being constant at~$O(\eps)$; this relies on the fact that any smooth perturbation of a two dimensional round two-sphere or flat space can be written as a scalar (or Weyl) perturbation of that space (which is not true for higher  dimensions, where non-trivial tensor perturbations can leave the Ricci scalar invariant at~$O(\eps)$), and we will discuss it more explicitly in Sections~\ref{subsec:PTflat} and~\ref{subsec:PTsphere}.  Now, the Bianchi identity implies that the trace-free bulk optical Ricci tensor obeys the condition
\be
\overline{\grad}^I \left( \bar{R}_{IJ} - \frac{1}{3} \bar{R} \bar{g}_{IJ} \right) = \frac{1}{6} \overline{\grad}_J \bar{R} \, ;
\ee
moreover, the boundary value of the optical Ricci scalar~$\bar{R}$ is (up to a positive constant) the value of the boundary Ricci scalar~$R$.  But since any nontrivial deformation of the boundary metric must yield a non-constant~$R$ at~$O(\eps)$, we conclude from the above equation that under any such deformation, the trace-free bulk optical Ricci tensor is nonvanishing at~$O(\eps)$.  Comparing with~\eqref{eq:EvacRsquared}, we thus conclude that under \emph{any} nontrivial deformation of the boundary geometry,~$\Evac[g + \eps h]$ becomes \textit{strictly negative} at~$O(\eps^2)$ (and vanishes for~$\eps = 0$).  But as shown above, the~$O(\eps^2)$ variation to~$\Evac$ is determined from universal CFT correlators, and thus we conclude that this property must hold in any (unitary) CFT.  This verifies that the bound~\eqref{eq:Evacbound} extends to all CFTs for~$\Sigma$ in the neighborhood of flat space or a round sphere.

Next, let us consider a more general weighting function~$f(x)$. Again $R$ is constant on flat space or the round sphere and is non-constant at order~$O(\eps)$ for any deformation away from these.  We require the weighting function~$f(R)$ to vary nontrivially at~$O(\eps^0)$ to obtain a nontrivial family of bounds; we therefore replace the weighting function with~$f(R/\eps)$ (so that variations of~$f$ are~$O(\eps^0)$).  Then following \cite{FisHic16} to derive~\eqref{eq:frhobound}, we have
\begin{subequations}
\bea
\int_\Sigma f(R/\eps) \rho_\mathrm{vac} &= \int_{\overline{\Sigma}} \overline{\grad}^I \left( \bar{f}(\bar{R}/\eps) \overline{\grad}_I \bar{R} \right), \\
		&= \int_{\overline{\Sigma}} \left[ \bar{f}(\bar{R}/\eps) \overline{\grad}^2 \bar{R} + \frac{1}{\eps} \bar{f}'(\bar{R}/\eps) \left(\overline{\grad} \bar{R}\right)^2 \right], \\
		&= \int_{\overline{\Sigma}} \left[-3\bar{f}(\bar{R}/\eps) \left(\bar{R}_{IJ} - \frac{1}{3} \bar{R} \bar{g}_{IJ}\right)^2 + \frac{1}{\eps} \bar{f}'(\bar{R}/\eps) \left(\overline{\grad} \bar{R}\right)^2 \right]. \label{eq:fboundstep}
\eea
\end{subequations}
We have already established that the first term on the right-hand side is~$O(\eps^2)$.  On the other hand, the second term is~$O(\eps)$, since~$\bar{f}'(\bar{R}/\eps)$ is~$O(\eps^0)$ and~$\overline{\grad} _I \bar{R}$ vanishes to order~$O(\eps)$ if and only if the Ricci scalar $R$ is constant at order~$O(\eps)$.  Thus (since~$\bar{f}'(x) < 0$ somewhere for nontrivial~$f$), for \emph{any} nontrivial perturbation taking $g$ away from being flat or round (and hence away from a constant $R$ at~$O(\eps)$) the integral of the second term is \textit{strictly negative} at~$O(\eps)$.  But since we argued above that the~$O(\eps)$ variation in~$\rho_\mathrm{vac}$ is universal, this property must hold perturbatively in any CFT as well.  This verifies that the bound~\eqref{eq:frhobound} extends perturbatively to all CFTs for a smooth~$f(x)$, i.e.~it holds in a neighborhood of the round sphere or flat space metrics on $\Sigma$.  In the limit that~$f(x)$ becomes a step function, one recovers the bound~\eqref{eq:rhobound} with the caveat that the ``step'' in~$f(R)$ cannot lie at a value of $R$ where  $\nabla R = O(\eps^2)$ everywhere on that level set of $R$ in $\Sigma$\footnote{This very non-generic situation (recall generically we will have $\nabla R = O(\eps)$) could potentially result in the second term in the integral~\eqref{eq:fboundstep} vanishing at~$O(\eps)$ (preventing us from invoking the same universality arguments) if this level set of $R$ extends in the bulk to a two-dimensional level set of $\bar{R}$ which also satisfies $\bar{\nabla} \bar{R} = O(\eps^2)$ on its entirety. We suspect it is not possible to find an~$\bar{R}$ surface with this property, but currently cannot exclude it.}.  Thus modulo this minor caveat, we perturbatively reproduce the bound~\eqref{eq:rhobound} as well.

Finally let us turn to the holographic result that~$\rho_{\mathrm{vac}}$ must be positive somewhere on~$\Sigma$ for~$\Sigma$ genus zero or one. The perturbative extension of this to any CFT for a perturbation of the sphere or flat space follows elegantly from the discussion above. We have seen that for any perturbation away from flat space or the round sphere we have $\Evac = \int_\Sigma  \rho_\mathrm{vac} \sim O(\epsilon^2)$. Choosing $r = \left( \min{R} + \max{R} \right)/2$ we also have $\int_{R \le r} \rho_\mathrm{vac} < 0$ and $\int_{R \le r} \rho_\mathrm{vac} \sim O(\eps)$ (modulo the caveat mentioned above, in which case one would simply take a different $r$). Clearly this implies $\int_{R \ge r} \rho_\mathrm{vac} > 0$ and $\int_{R \ge r} \rho_\mathrm{vac} \sim O(\eps)$ which in turn requires that $\rho_\mathrm{vac}$ be positive somewhere in the region~$R \ge r$ of~$\Sigma$.

While powerful, these arguments were indirect in the sense that they never explicitly compute the corrections to~$\Evac$ or~$\rho_\mathrm{vac}$.  As a check, let us now show explicitly how~$\Evac$ and~$\rho_\mathrm{vac}$ vary under perturbations.  We perform these computations by using a holographic dual to evaluate the 1-point function $\langle T_{AB} \rangle_{g + \eps h}$, yielding $\rho$ and $t_{ij}$ and then via~\eqref{eq:secondvarE} we obtain~$\Evac$. We note that one could equivalently compute directly in CFT language using the universal form of the two-point function~\eqref{eq:TTuniversal} and explicitly computing the integral in~\eqref{eq:lintijrho}. An advantage of our approach is that the UV divergences are simply regulated using the holographic renormalization prescription, and there is no IR divergence which one would encounter when performing the integral over $\tau$ in~\eqref{eq:lintijrho} -- this has already been conveniently taken care of in the discussion leading to~\eqref{eq:secondvarE}. Of course since $t_{ij}$ in~\eqref{eq:lintijrho} has a universal form, however one computes it one must obtain the same result, and in the end simplicity is essentially a matter of taste.

\subsection{Deforming a CFT on Static $\mathbb{R}^2$}
\label{subsec:PTflat}

Let us consider ``static'' metric perturbations to Euclidean flat three-dimensional space,
\be
ds^2 = d\tau^2 + \left(\delta_{ij} + \epsilon  h_{ij}\right) dx^i dx^j,
\ee
where the perturbation~$h_{ij}$ on~$\Sigma$ is~$\tau$-independent.  In general,~$h_{ij}$ can be decomposed into scalar, vector, and tensor perturbations, but in two dimensions none of the tensor perturbations are independent, and the vector perturbations just correspond to diffeomorphisms.  Thus the only nontrivial perturbations are scalars, which we decompose into Fourier modes:
\be
\label{eq:flatspaceperturbation}
h_{ij} = \chi(x^k) \delta_{ij}, \quad \chi(x^i) = \int d^2k \, \chi(k_i) e^{-i k_j x^j},
\ee
where reality of~$h_{ij}$ implies~$\chi(-k_i) = \chi^*(k_i)$.  Note that the~$\chi(k = 0)$ mode corresponds to an overall rescaling of~$\Sigma$, which in turn corresponds to an overall Weyl rescaling of the full geometry~$\rnum \times \Sigma$ leaving it locally conformally flat.  Such a rescaling is trivial, and therefore we are uninterested in these~$k = 0$ modes.  However, the~$k \neq 0$ modes are all non-trivial, deforming the geometric away from local flatness so that the Ricci scalar~$R$ of $\Sigma$ is non-zero at order~$O(\eps)$.

Now, we write the bulk solution (which is assumed static) as
\begin{multline}
ds^2_\mathrm{bulk} = \frac{l^2}{z^2} \left[ d\tau^2  + \delta_{ij} \left( 1 + \eps A(z, x^k) \right) dx^i \, dx^j \right. \\ \left. + \left(1 + \eps B(z, x^k) \right)  dz^2 + 2 \eps \partial_i C(z, x^k) dx^i dz \right] + O(\eps^2);
\end{multline}
clearly the boundary perturbation is~$\chi(x^i) = A(0,x^i)$.  Then for a single Fourier mode we write~$A = e^{- i k_j x^j} a(z)$,~$B = e^{- i k_j x^j} b(z)$, and $C = e^{- i k_j x^j} c(z)$ and find that the bulk equations of motion at~$O(\eps)$ reduce to
\begin{subequations}
\be
\partial_z^2 a - \frac{2}{z} \partial_z a - k^2 a = 0,
\ee
\be
b(z) =  - \frac{z}{2} a'(z), \quad c(z) = - \frac{k^2 z \,a(z) + a'(z)}{4 k^2}.
\ee
\end{subequations}
For appropriate boundary conditions at the Poincar\'e horizon, the first is solved by
\be
a(z) = a_0 e^{-k z} \left( 1 + k z \right),
\ee
so we identify~$\chi(k_i) = a_0(k_i)$.  The stress tensor one-point function may be extracted in the standard way~\cite{BalKra99,deHSol00} by converting to Fefferman-Graham coordinates~$(Z,X^i)$ given by
\begin{subequations}
\bea
z &= Z \left( 1 + \epsilon a_0 \left( -\frac{k^2}{8} Z^2  + \frac{1}{12} k^3 Z^3 + O(Z^4) \right) e^{- i k_j x^j} \right) + O(\eps^2), \\
x^i &= X^i  + \epsilon a_0 \left( - \frac{1}{12} i k\, k^i Z^3 + O(Z^4) \right) e^{- i k_j x^j} + O(\eps^2),
\eea
\end{subequations}
from which we obtain the spatial stress tensor
\be
\frac{16\pi G_N}{l^2} t_{ij}[\delta + \eps h] = \frac{\eps}{2} \, a_0 e^{-i k_j x^j} k \left(k^2 \delta_{ij} - k_i k_j\right).
\ee
Thus for the general perturbation~\eqref{eq:flatspaceperturbation} we find, using the identification~$a_0(k_i) = \chi(k_i)$ and~\eqref{eq:cTgrav}, that the boundary stress tensor is
\be
\label{eq:flatstresstensor}
t_{ij}[\delta + \eps h] = \frac{\eps}{2} \frac{\pi^2 c_T}{48} \int d^2 k \, \chi(k_k) k(k^2 \delta_{ij} - k_i k_j) e^{-i k_j x^j},
\ee
and therefore from~\eqref{eq:secondvarE} we find that
\be
\label{eq:2dflatvariation}
\Evac[\delta +  \eps h] = \Evac[\delta] - \eps^2\, \frac{\pi^4 c_T }{96} \int d^2k  \left| \chi(k) \right|^2  k^3 + O(\eps^3).
\ee
This immediately implies that any non-trivial perturbation (ie.~$k \neq 0$) decreases~$\Evac$ to leading nontrivial order in~$\eps$.
Hence we see precise agreement with the less constructive argument in Section~\ref{subsec:pertbounds} which implied such deformations must decrease~$\Evac$ at~$O(\eps^2)$.

As a final point, let us obtain a more explicit expression for~$\rho_\mathrm{vac}$.  For a single Fourier mode~$h_{ij}(x) = \chi(k_k) e^{-i k_j x^j} \delta_{ij}$, the boundary Ricci scalar~$R$ and energy density (computed from~\eqref{eq:linrho}) are
\begin{subequations}
\bea
\label{eq:LorRiccirho}
R(x) &= \eps \chi(k_i) k^2 e^{-i {k}_j x^j}+ O(\eps^2), \\
\rho_\mathrm{vac}(x) &= \eps \, \frac{\pi^2 c_T}{96} \chi(k_i)   k^3 e^{- i k_j x^j} + O(\eps^2).
\eea
\end{subequations}
We see  explicitly that indeed there exist regions on~$\Sigma$ where $\rho_\mathrm{vac} > 0$ for any single mode.  
Interestingly, in this language it is not at all obvious how the bound $\int f(R) \rho_\mathrm{vac} \le 0$ arises. With the explicit expressions above, directly obtaining the bound is a non-trivial challenge requiring integration over modes to form a deformation of localized support. Nevertheless, the holographic calculations of Section~\ref{subsec:pertbounds} guarantee the bound must hold.  This is a nice illustration that robust bounds derived in the holographic context may be highly non-trivial, and indeed non-obvious, from the purely CFT perspective, showing the power of holography to give potential new statements for CFTs in this curved spacetime context.

\subsection{Deforming a CFT on the Static $S^2$}
\label{subsec:PTsphere}

Next, let us repeat the above analysis for perturbations of a round sphere.  The space~$\rnum \times S^2$ can be obtained via a Weyl rescaling of three-dimensional Euclidean space by taking a Weyl factor of~$\omega(x) = 1/|x|$, with~$|x|$ the usual Cartesian distance.  Specifically, under the coordinate transformations
\be
x^1 = e^\tau \sin{\theta} \cos{\phi}, \quad x^2 = e^\tau \sin{\theta} \sin{\phi}, \quad x^3 = e^\tau \cos{\theta},
\ee
the Weyl rescaled metric~$G_{AB} = \omega^2(x) \delta_{AB}$ takes the form
\be
ds^2 = d\tau^2 + \Omega_{ij} d\theta^i d\theta^j,
\ee
where~$\Omega_{ij}d\theta^i d\theta^j = d\theta^2 + \sin^2{\theta} \, d\phi^2$ is the usual metric on the round~$S^2$.  

Since~$\rnum \times S^2$ is locally conformal to flat Euclidean space, the results of Section~\ref{subsec:partfuncproperties} apply.  Let us therefore consider a perturbation of the round sphere; as above, the only nontrivial independent perturbations are scalars, so we consider a perturbation to the sphere of the form
\be
\label{eq:twospherepert}
ds^2_{\Omega + \eps h} = (1 + \eps \chi(\theta^i)) \Omega_{ij} d\theta^i d\theta^j.
\ee
Again, it is natural to decompose~$\chi$ into spherical harmonics~$Y_{\ell m}$:
\be
\label{eq:spherechimodes}
\chi(\theta^i) = \sum_{\ell m} \chi_{\ell m} Y_{\ell m}(\theta^i).
\ee
Note that the only nontrivial modes have~$\ell > 1$, for the~$\ell = 0$ mode just corresponds to a rigid rescaling of the sphere (which leaves the full conformal geometry~$M$ invariant), while the three~$\ell = 1$ modes just correspond to infinitesimal diffeomorphisms corresponding to rigid rotations of the sphere.  For~$\ell > 1$ these scalar modes deform the Ricci scalar $R$ to be non-constant at order~$O(\eps)$.

Writing the bulk metric as
\begin{multline}
\label{eq:spherebulk}
ds^2_\mathrm{bulk} = \frac{l^2}{z^2} \left[ \left( 1 + z^2 \right) d\tau^2  + \Omega_{ij} \left( 1 + \eps A(z, \theta^k) \right) d\theta^i d\theta^j \phantom{\frac{dz^2}{1+z^2}} \right. \\ \left. + \left( 1 + \eps B(z, \theta^k) \right)   \frac{dz^2}{1+z^2} + 2 \eps \, \partial_i C(z, \theta^k) d\theta^i dz \right],
\end{multline}
the boundary perturbation is~$\chi(\theta^i) = A(0,\theta^i)$.  Now we decompose the perturbation into spherical harmonics on the~$S^2$ and consider a single such mode: $A(z, \theta^k)  = Y_{\ell m}(\theta^k) a(z)$, $B(z, \theta^k)  = Y_{\ell m}(\theta^k) b(z)$ and $C(z, \theta^k)  = Y_{\ell m}(\theta^k) c(z)$.  The bulk equations then yield
\begin{subequations}
\be
\partial_z^2 b - \frac{4}{z} \partial_z b + \left( \frac{6}{z^2} - \frac{\ell(\ell+1)}{1+z^2} \right) b = 0, \label{eq:bequation}
\ee
with
\bea
a &= \frac{1}{(\ell-1)(\ell+2)} \left[ \left(\frac{6}{z^2} + 4 \right) b - \left(\frac{2}{z}+z \right) \partial_z b \right], \\
c &= \frac{1}{(\ell-1)(\ell+2)} \left[ - \frac{ ( 1 + 2 z^2)}{ z (1+z^2)} b + \frac{1}{2} \partial_z b \right].
\eea
\end{subequations}
The solution to~\eqref{eq:bequation} which is regular in the bulk is
\begin{subequations}
\begin{multline}
\label{eq:twospherefnb}
b = b_0 \frac{(\ell-1)(\ell+2)}{2}  \left( {}_2F_1\left[-\frac{\ell+1}{2}, \frac{\ell}{2}  , \frac{1}{2} , -z^2 \right] z^2  \right. \\ \left. + 2 p_\ell \, {}_2F_1\left[ -\frac{\ell}{2} , \frac{\ell+1}{2}  , \frac{3}{2} , -z^2 \right] z^3 \right),
\end{multline}
with
\be
p_\ell \equiv \frac{\ell+1}{\ell} \left(\frac{\Gamma\left[(\ell+1)/2\right]}{\Gamma\left[\ell/2\right]}\right)^2,
\ee
\end{subequations}
in terms of which the boundary perturbation is~$\chi(\theta^i) = b_0 Y_{\ell m}$.  To obtain the stress tensor, we again change to Fefferman-Graham coordinates~$(Z,\Theta^i)$ via
\begin{subequations}
\bea
z &= Z \left( 1 + \frac{1}{4} Z^2 + \eps b_0 (\ell + 2)(\ell-1) \left( - \frac{1}{8}   Z^2  + \frac{1}{6}  p_\ell Z^3  \right) Y_{\ell m} + O(Z^4) \right), \\
\theta^i &= \Theta^i  + \eps  \frac{b_0}{6} p_\ell  \Omega^{ij} \grad_j Y_{\ell m} Z^3 + O(Z^4),
\eea
\end{subequations}
from which we finally obtain
\be
\frac{16\pi G_N}{l^2} t_{ij}[\Omega + \eps h] = \eps \, b_0 p_\ell \left[ \left( \ell(\ell+1) - 1 \right) \Omega_{ij} +\grad_i \grad_j  \right] Y_{\ell m}, 
\ee
with~$\grad_i$ the gradient on the~$S^2$.  Thus under an arbitrary perturbation~\eqref{eq:spherechimodes}, we have
\be
t_{ij}[\Omega + \eps h] = \eps \, \frac{\pi^2 c_T}{48}  \sum_{\ell m} \chi_{\ell m} p_\ell \left[ \left( \ell(\ell+1) - 1 \right) \Omega_{ij} +\grad_i \grad_j \right] Y_{\ell m}.
\ee
Then using~\eqref{eq:secondvarE}, orthogonality of the spherical harmonics, and reality of the metric perturbation, we find
\be
\Evac[\Omega + \eps h] = \Evac[\Omega] - \frac{\eps^2}{4} \frac{\pi^2 c_T}{48} \sum_{\ell m} p_\ell (\ell-1)(\ell+2) \left| \chi_{\ell m}  \right|^2 + O(\eps^3).
\ee
It is easy to show that~$p_\ell$ is positive for all~$\ell > 0$ and vanishes for~$\ell = 0$; thus this perturbation vanishes only for~$\ell = 0$ and~$\ell = 1$.  But these are precisely the trivial modes we excluded, and thus we find that any nontrivial deformation of the sphere (which perturbs $R$ from being constant at order~$O(\eps)$)
does indeed decrease~$\Evac$ at leading nontrivial order~$O(\eps^2)$, again in precise agreement with the non-constructive argument in Section~\ref{subsec:pertbounds}.

Finally, let us again examine the behavior of~$\rho_\mathrm{vac}$.  For a single mode, the Ricci scalar and energy density (computed from~\eqref{eq:linrho}) of the perturbed sphere are given by
\begin{subequations}
\bea
\label{eq:sphereRicci}
R &= 2 + \eps \, \chi_{\ell m} (\ell-1)(\ell+2) Y_{\ell m} + O(\eps^2) \equiv 2 + \eps \, \delta R + O(\eps^2), \\
\rho_\mathrm{vac} &= \eps \, \frac{\pi^2 c_T}{48} \chi_{\ell m} p(\ell) (\ell-1)(\ell+2)Y_{\ell m} + O(\eps^2).
\eea
\end{subequations}
Again we see regions on~$\Sigma$ where $\rho_\mathrm{vac} > 0$ for any single mode.  For a single mode we also have~$\int_{S^2} \delta R = 0$, and thus~$\int_{R \leq r} \delta R \leq 0$ for any~$r$.  But since~$\delta R$ is proportional to~$\rho_\mathrm{vac}$, we again recover
\be
\int_{R \leq r} \rho_\mathrm{vac} \leq 0
\ee
for all~$r$.  As for flat space, the gravitational calculations of Section~\ref{subsec:pertbounds} require that this bound hold for any sum of modes although as we emphasized before, this condition is far from obvious in this language.

\subsection{Partition Functions on More General Euclidean Geometries}

So far we have focused exclusively on the behavior of the vacuum energy in conformal geometries~$\rnum \times \Sigma$ when~$\Sigma$ is a perturbative deformation of flat space or a round sphere.  But since~$\Evac$ is closely related to the partition function~$Z$, we may wonder whether these perturbative results lead to more general conditions than the ones we have presented so far.  This is indeed the case: on Euclidean geometries that are small deformations of flat space or the three-sphere, it is possible to show that~$\ln Z$ must \textit{increase} perturbatively.  Though not directly relevant to our purposes, we present these arguments in Appendices~\ref{app:flatpert} and~\ref{app:spherepert}.  In particular, for perturbations of flat Euclidean space we perform the calculation both holographically and in purely field theoretic terms to show explicitly that they match (as they must, since as we have shown the quadratic perturbation to~$\ln Z$ is given in terms of correlators with universal behavior).

This result is interesting in light of the fact that the behavior of~$\ln Z$ for CFTs on deformed spheres has received recent attention in~\cite{Bobev:2016sap,BobBue17}, motivated in part by the F-theorem \cite{JafKle11,Klebanov:2011gs}, where specific CFTs (holographic or free) where solved for a specific squashing deformation of the round sphere geometry non-perturbatively in $\eps$. The results in Appendix~\ref{app:spherepert} may be thought of as complementary to that work, describing the universal behaviour for general deformations of the 3-sphere at leading perturbative order in~$\eps$ for any CFT, rather than the specific squashing deformation to all orders in~$\eps$ for specific CFTs. 

Another fascinating potential application of Appendix~\ref{app:spherepert} (and the driving motivation for the work~\cite{Bobev:2016sap}) is that the behaviour of~$Z^{-1}$ for deformations of Euclidean 3-spheres has been argued to control the amplitude of the wavefunction of the universe for inhomogeneous final states \cite{Hertog:2011ky} in a holographic version of the no-boundary proposal \cite{HarHaw83} (see also \cite{McFadden:2009fg} for related ideas). Understanding that the homogeneous closed universe is preferred in this measure over an inhomogeneous closed one is precisely the same issue as whether $\ln Z$ is minimized for the round sphere --  Appendix~\ref{app:spherepert} shows this is indeed true perturbatively independent of the details of the CFT controlling this wavefunction providing it is unitary (so that $c_T > 0$).  Likewise, for a non-unitary CFT with~$c_T < 0$ the same perturbative calculation would indicate the opposite, namely an inhomogeneous universe would have a greater amplitude for its wave function than a homogeneous one.

Incidentally, one might think that the full Euclidean  perturbations of flat space in Appendix~\ref{app:flatpert} would provide the quickest route to establish \eqref{eq:2dflatvariation} for~$\rnum \times \Sigma$ with~$\Sigma$ perturbed flat space. This latter case can, after all, be thought of as a static perturbation of flat three-space.  However, such static perturbations are not normalizable (since they do not decay in the~$\tau$-direction). It would not be too difficult to control this divergence, but the holographic computation we performed in Section~\ref{subsec:PTflat} takes care of it automatically and was thus more convenient for our purposes.

%================================= Global Bounds =================================

\section{Beyond the Perturbative Regime}
\label{sec:nonpert}

In the previous section, we have shown using general CFT arguments that the holographic results on the vacuum energy~$\Evac$ and energy density~$\rho_\mathrm{vac}$ outlined in Section~\ref{sec:holog} are reproduced perturbatively in any CFT.  That is, we have shown that~(i)~$\Evac$ (thought of as a function of the metric on the space~$\Sigma$) achieves a local maximum when~$\Sigma$ is the round sphere or flat space; and~(ii) that for perturbations of the round sphere or flat space,~$\rho_\mathrm{vac}$ is negative when integrated inside level sets of the Ricci scalar of~$\Sigma$.

Result~(i) implies that in the space of geometries~$\Sigma$, there exists a small neighborhood of the round sphere or flat space on which~$\Evac$ is negative.  For a CFT with a classical gravity holographic dual, it is in fact the case that~\textit{any} geometry on~$\Sigma$ yields a negative~$\Evac$.  A natural question therefore arises: is this also the case for general CFTs?  That is, do there exist \textit{any} geometries (of given topology) on which~$\Evac$ is positive?  In this section we will give a partial answer to this question by showing that the neighborhood of geometries on which~$\Evac$ is negative must extend to the boundary of the space of geometries on~$\Sigma$; that is, this neighborhood is not ``perturbatively small''.

\subsection{Properties of $\Evac$}
\label{subsec:lnZproperties}

In order to move away from the perturbative regime, we must first establish (or assume) certain properties of the vacuum energy~$\Evac$.  First, we must of course restrict ourselves to the space of (conformal) geometries on which the theory is well-defined; let us call this space of geometries~$\Mcal$.  Since we are restricting ourselves to geometries that are in the same conformal class as~$\rnum \times \Sigma$, we will use~$\Sigma$ to label these full conformal geometries.  Since we are considering the zero temperature limit, the Euclidean ultrastatic condition defines a preferred conformal frame up to an overall global scaling, so we may think of this labelling as unique up to rigid scaling; in other words, we will think of~$\Evac[\Sigma]$ as a functional on~$\Sigma \in \Mcal$ with invariance under rigid scaling of $\Sigma$.

We emphasize that not all regular geometries~$\Sigma$ necessarily yield a well-defined partition function, and therefore~$\Mcal$ is generically a proper subset of all possible regular geometries (up to scale).  To see this in a simple example, consider the free conformal scalar field~$\Phi$ discussed briefly in Section~\ref{subsec:holoscalars}.  Because the action is quadratic in~$\Phi$, the Euclidean path integral defining~$Z$ is Gaussian:
\be
Z[\Sigma] = \int \Dcal \Phi \, e^{-\int d\tau \int_\Sigma \Phi L \Phi},
\ee
where
\be
L \equiv -\partial_\tau^2 - \grad^2 + \frac{(d-2)}{4(d-1)} R,
\ee
and as usual~$\grad^2$ is the Laplacian of~$\Sigma$ and the volume element on~$\Sigma$ in the exponent is understood.  Now, when~$R$ is positive,~$L$ has only positive eigenvalues and therefore the Gaussian integral converges in every direction in the configuration space of~$\Phi$ (potentially up to a standard divergence in the zero-temperature limit associated to the infinite range of~$\tau$).  In such a case, the path integral is well-defined.  However, if~$R$ is somewhere negative, it's possible for~$L$ to obtain negative eigenvalues, corresponding to directions in the configuration space of~$\Phi$ in which the path integral diverges; in such a case,~$Z$ (and therefore the theory) is ill-defined.

Explicitly, consider the subset of the domain of the path integral consisting of~$\tau$-independent configurations~$\Phi$; on this subspace, the operator~$L$ reduces to the operator on the right-hand side of~\eqref{eq:freeconformalspectrum}.  But as shown in Section~\ref{subsec:holoscalars}, on the deformed sphere metric~\eqref{eq:deformedsphere} the lowest eigenvalue of this operator satisfies~$\omega^2_\mathrm{min} < 0$ as long as~$r > 15/4$, and therefore~$L$ must have at least one negative eigenvalue as well.  Thus even though the metric~\eqref{eq:deformedsphere} is perfectly regular, for~$r > 15/4$ there exists at least one direction in the configuration space of~$\Phi$ along which the Gaussian integral defining~$Z$ does not converge.  We thus have an example of a case where a regular geometry does not lead to a well-defined theory, and confirms our earlier statement that generically the space~$\Mcal$ is a proper subset of the space of regular geometries.

Indeed, from the holographic perspective this feature was briefly alluded to in Section~\ref{sec:holog}.  The bounds reviewed there relied implicitly on the existence of a bulk Einstein metric (possibly satisfying further restrictions), but it is unclear whether such metrics exist given an arbitrary $\Sigma$.  For example in~\cite{Anderson:2002xb} it was argued a four-dimensional bulk without horizons exists provided $\mathbb{R} \times \Sigma$ has positive scalar curvature in a particular conformal frame, but in the case of negative scalar curvature there are no existence results, and Witten's original argument~\cite{Wit98a} may suggest a bulk might not generally exist precisely due to the dual CFT likely being ill-defined as in our free conformal scalar example (see also~\cite{Witten:1999xp}). For such $\Sigma$ there may be no semiclassical bulk description of the holographic CFT, presumably indicating either that it is strongly quantum, or that it is not well-defined. Even if smooth bulk metrics do exist for some $\Sigma$, it is conceivable that the metric that defines the semiclassical saddle point may suffer from tachyonic instabilities, and if these resided in the universal gravity sector, they would render any holographic theory on $\Sigma$ ill-posed.  It may also be that despite the gravity dual appearing sound, there is a pathology of the theory associated with stringy physics beyond the gravitational description. While such questions are interesting we leave them as avenues for future study, and here only emphasize that even though we have holographic bounds that apparently apply for wide classes of $\Sigma$, this is contingent on the existence of bulk metrics, and also that the resulting metric defines a stable vacuum state not just within gravity but within a full stringy description.

Next, we will need some smoothness assumptions on~$\Evac[\Sigma]$.  Recall that as discussed in Section~\ref{sec:PT},~$\Evac[\Sigma]$ is defined as an appropriate zero-temperature limit of~$-\ln Z[\Sigma]$.  Now, in general we expect~$Z[\Sigma]$ to be continuous; however, it is possible for~$Z[\Sigma]$ to not be everywhere differentiable, corresponding to first-order phase transitions as the geometry is varied.  Thus~$\Evac[\Sigma]$ should also be continuous, but may not be differentiable at first-order phase transitions.  Nevertheless, it is reasonable to expect that~$\Evac[\Sigma]$ should be differentiable at any local minima; that is, none of its local minima should be ``cusp-like''.  The intuition for this expectation is perhaps clearest in the Lorentzian picture.  In that case, the phase transitions identify locations in~$\Mcal$ where the state with the lowest free energy changes (and thus right at the phase transition, the vacuum is degenerate).  Thus any cusps in~$\Evac[\Sigma]$ (which \textit{is} the free energy at zero temperature) indicate the presence of other states with a free energy different (and necessarily higher) than that of the vacuum state for $\Sigma$ near the phase transition geometry.  
But cusp-like local minima of~$\Evac[\Sigma]$ would indicate the presence of states near the transition geometry with free energy \textit{lower} than that of the vacuum at the transition geometry (essentially by ``continuing through'' the cusp),  contradicting the condition that the free energy be minimized by the vacuum state. Therefore their existence must be excluded.  While this argument is not precise, it motivates the following assumption: we will assume that \textit{all local minima of~$\Evac[\Sigma]$ in~$\Mcal$ are critical points}.

This assumption has an important implication, which we will exploit below.  From \eqref{eq:secondvarElin}, the spatial vacuum stress tensor can be obtained from a functional derivative of~$\Evac[\Sigma]$; thus the spatial vacuum stress tensor must vanish at any critical point of~$\Evac[\Sigma]$.  Moreover, by the tracelessness of the full stress tensor, a vanishing spatial stress tensor implies a vanishing energy density, and therefore the energy density vanishes at critical points of~$\Evac[\Sigma]$ as well.  But since~$\Evac$ can be expressed as an integral over this energy density, we conclude that~$\Evac$ must vanish at any of its critical points.  Combined with the above assumption, we infer that~$\Evac$ must vanish at any of its local minima.  We may now exploit this result to extend the perturbative results for~$\Evac$ away from flat space and the round sphere.

\subsection{Non-Perturbative Bounds}
\label{subsec:nonpert}

Assume there exists some geometry~$\overline{\Sigma}$ with negative vacuum energy, and consider the space~$\Xcal_{\overline{\Sigma}}$ of geometries which are connected (in~$\Mcal$) to~$\overline{\Sigma}$ and on which~$\Evac[\Sigma]$ is nonpositive:
\be
\Xcal_{\overline{\Sigma}} \equiv \left\{ \Sigma \in \Mcal \middle| \Evac[\Sigma] \leq 0 \mbox{ and } \Sigma \mbox{ is connected to } \overline{\Sigma}\right\}.
\ee
Roughly speaking,~$\Xcal_{\overline{\Sigma}}$ is the space of all conformal geometries with nonpositive energy that are smoothly deformable to~$\overline{\Sigma}$.  Our purpose is to show that~$\Xcal_{\overline{\Sigma}}$ is not bounded in~$\Mcal$; that is, that it must extend to the boundary of~$\Mcal$.

We will proceed by contradiction.  Assume that~$\Xcal_{\overline{\Sigma}}$ is in fact bounded; because~$\Evac[\Sigma]$ is continuous, it must be that~$\Evac[\partial \Xcal_{\overline{\Sigma}}] = 0$; thus~$\Evac$ must achieve its minimum value in~$\Xcal_{\overline{\Sigma}}$ in the interior of~$\Xcal_{\overline{\Sigma}}$, and therefore this minimum must be a local minimum.  But such a local minimum cannot exist, since~$\Evac$ would need to be negative there, whereas we established above that~$\Evac$ must vanish at all of its local minima.  We therefore have a contradiction and conclude that~$\Xcal_{\overline{\Sigma}}$ cannot be bounded in~$\Mcal$.

\begin{figure}
\centering
\includegraphics[width=0.35\textwidth,page=1]{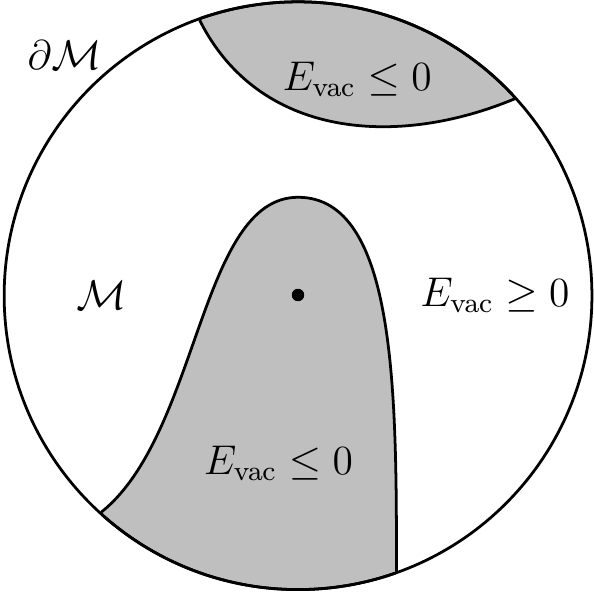}
\caption{A schematic depiction of the partition of~$\Mcal$ into regions of nonpositive and nonnegative~$\Evac$.  Any regions of nonpositive~$\Evac$ (gray) must extend to the boundary of~$\Mcal$ (i.e.~there can be no ``islands'' of nonpositive~$\Evac$).  The black dot represents the round sphere or flat space; these both lie within a region of nonpositive~$\Evac$.}
\label{fig:Mspace}
\end{figure}

Thus any regions of nonpositive~$\Evac$ must extend to the boundary of~$\Mcal$; that is, a partition of~$\Mcal$ into regions of nonpositive~$\Evac$ must look schematically as shown in Figure~\ref{fig:Mspace}.  Interestingly, because we have only assumed that local minima of~$\Evac[\Sigma]$ are critical points (and in particular, we have made no claim regarding its local maxima), we cannot invert this argument to obtain any conclusions about regions with non\textit{negative}~$\Evac$\footnote{Though it is worth noting that if we were to assume that all local extrema of~$\Evac[\Sigma]$ were critical points, we would indeed be able to invert the argument to conclude that any regions of nonnegative~$\Evac$ must also extend to the boundary of~$\Mcal$.}.  

Of particular interest is the region containing the round two-sphere~$S^2$ or flat space~$\mathbb{R}^2$ (depending on the topology of the space~$\Sigma$ under consideration).  By the perturbative calculations of Section~\ref{sec:PT}, flat space and the round sphere (modulo scale) belong to a region of nonpositive~$\Evac$.  We have thus obtained the promised global generalization of the results of Section~\ref{sec:PT}: the neighborhood of the round~$S^2$ or flat~$\mathbb{R}^2$ with nonpositive vacuum energy extends to the boundary of the space of conformal geometries on which the theory is well-defined.  In other words, the round sphere or flat space can be continuously deformed to a geometry on which the theory is ill-defined while maintaining a nonpositive vacuum energy.  Moreover, any nontrivial perturbation of the round~$S^2$ or flat~$\mathbb{R}^2$ generically initiates a trajectory of steepest descent leading to the boundary of~$\Mcal$.

This result makes it natural to wonder whether~$\Evac$ is always non-positive over \emph{all} of $\mathcal{M}$ for these topologies.  Indeed, this is precisely what the holographic bound~\eqref{eq:Evacbound} implies for holographic CFTs (modulo caveats regarding the existence of a bulk dual).  Unfortunately, we have no convincing argument that allows us to make such a general statement for all CFTs, and it may well be that for some CFTs more complicated pictures, such as the schematic one depicted in Figure~\ref{fig:Mspace}, occur.  Perhaps even if it is not generally true, one might hope it may be the case for many classes of $(2+1)$-dimensional CFTs beyond just holographic ones.  Conceivably this could be tested by employing numerical methods like those used in~\cite{BobBue17}; it is clear that the behavior of $\Evac$ near the boundary of $\Mcal$ is crucial in determining whether $\Evac$ can anywhere be positive, and perhaps near-degenerate metrics may provide a possible avenue to understand global properties of $\Evac$.

Finally, let us comment on potential generalizations to~$\ln Z$.  The above argument made crucial use of properties that~$\Evac$ inherits from~$\ln Z$; since the perturbative results of Appendices~\ref{app:flatpert} and~\ref{app:spherepert} find that~$\ln Z$ is locally minimized on the three-sphere and flat Euclidean space, it is natural to ask whether there are analogous global statements we can make for~$\ln Z$.  The answer appears to be no.  A crucial ingredient in the argument above was that~$\Evac$ must vanish at critical points of~$\ln Z$, but there is no analogous statement constraining the behavior of~$\ln Z$ at these points.  Without such a constraint, the argument does directly not extend to~$\ln Z$. Nonetheless, if the space of geometries on which~$\ln Z$ exceeds its value on the round three-sphere or flat Euclidean space did not extend to the boundary of~$\Mcal$, it would imply the existence of an additional local extremum of~$\ln Z$.  While the existence of such a geometry is possible, it seems unlikely as it would need to have vanishing stress tensor.  This intuition leads us to believe that it is likely that just as for~$\Evac$, there are directions along which~$\ln Z$ increases from its local minimum on the round three-sphere or flat Euclidean space all the way to the boundary of~$\Mcal$.

Given this tentative result, one might also wonder whether our perturbative analysis of~$\ln Z$ about the round Euclidean three-sphere in Appendix~\ref{app:spherepert} suggests $\ln Z$ must always be greater than its value on the round sphere.
This cannot be, as the recent study of~\cite{BobBue17} showed explicitly that~$\ln Z$ for a free Dirac fermion under a sufficiently large squashing deformation attains a value less than that on the round sphere.  However, given that a potential global bound on~$\ln Z$ is violated by the free Dirac Fermion, this observation offers an interesting testing ground for the putative global bound on~$\Evac$: namely, is~$\Evac$ always nonpositive for the free Dirac fermion when~$\Sigma$ is a squashed sphere?

\subsection{Dissipative Dynamics}
\label{subsec:dynamics}

The above global result on~$\Evac$ allows us to make an interesting observation regarding the dynamics of any system governed by~$\Evac$.  Consider a CFT living on~$\rnum \times \Sigma$ but where now~$\Sigma$ is permitted to evolve dynamically (for instance, we may imagine a CFT living on a dynamical membrane embedded in some ambient spacetime).  Next, assume that the dynamics of this system is driven solely by the vacuum energy of the CFT (so for the dynamical membrane example the tension would have to be negligible), and is dissipative in the sense that this vacuum energy decreases through the evolution.   In such a configuration, the system will evolve in a direction that minimizes~$\Evac$.  Since~$\Sigma$ is now permitted to evolve, our results above imply that, as long as~$\Evac$ is initially negative (by, for instance, taking~$\Sigma$ to be a small perturbation of a sphere), this evolution will become singular when~$\Sigma$ reaches the boundary of the space~$\Mcal$.  This endpoint may correspond to the geometry~$\Sigma$ becoming genuinely singular, or to some other pathology in the theory on~$\Sigma$.  

This observation leads to a natural question: is there a obvious way to encode such a dynamics on~$\Mcal$?  A simple option is to  let the geometry of~$\Sigma$ vary as the spatial stress tensor:
\be
\label{eq:Tflow}
\frac{d g_{ij}}{d\lambda} = -D \, t_{ij}[g] + \grad_{(i} \xi_{j)},
\ee
where~$D$ is an arbitrary coefficient and~$\lambda$ is the ``dynamical time'' along the flow of geometries.  Note that~$\xi^i$ is an arbitrary vector field (which may depend on~$\lambda$) which allows for flow-dependent diffeomorphisms; it encodes the fact that we are only interested in flowing the geometry and not (the components of) the metric.  We will call this flow the $T$-flow\footnote{More generally, we could define a flow over the full conformal~$3$-d geometry~$G_{AB}$ as~$dG_{AB}/d\lambda = -D\ev{T_{AB}}_g + \grad_{(A}\xi_{B)} + f G_{AB}$, where like~$\xi^A$,~$f$ is an arbitrary function (which also may depend on~$\lambda$) which allows for flow-dependent Weyl transformations.  Such a generalization is unnecessary to our present purposes and we will not discuss it further here, except to note that~\eqref{eq:Tflow} can be thought of as a special case of such a generalized flow.}.
\emph{Naively} this flow is indeed dissipative for $D < 0$, as it is then the natural gradient descent of $\Evac$. This follows from the fact that from~\eqref{eq:secondvarElin},~$t_{ij}$ is defined by a functional derivative of~$\Evac$ with respect to $g_{ij}$, and hence the flow (for either sign of $D$) is a gradient flow of $\Evac$:
\be
\frac{d\Evac}{d\lambda} = \int_\Sigma \frac{1}{\sqrt{g}} \, \frac{\delta \Evac}{\delta g_{ij}} \frac{dg_{ij}}{d\lambda} = \frac{D}{2} \int_\Sigma \left|t_{ij}[g] \right|^2 
\ee
where the diffeomorphism term~$\grad_{(i}\xi_{j)}$ vanishes via an integration by parts and conservation and staticity of the stress tensor. Hence for~$D < 0$ ($D > 0$) this is a gradient descent (ascent) of~$\Evac$.

Interestingly, this dissipative~$T$-flow dynamics with $D < 0$ is ill-posed, and the $T$-flow only exists as an (unphysical, at least expecting dynamics to be dissipative) gradient \textit{ascent} of~$\Evac$. This can be seen by a perturbative analysis. The idea is that at sufficiently small scales, the flow can be approximated by its behavior near flat space.  As in Section~\ref{subsec:PTflat} we consider a scalar perturbation of the locally flat 2-dimensional space, 
\be
g_{ij}(\lambda;x^i) = \delta_{ij} + \eps \,\delta_{ij} \int d^2 k \, \chi(\lambda;k_i) e^{-i k_i x^i}, 
\ee
which may be chosen for all flow time applying a suitable diffeomorphism~$\xi_i$. The stress tensor is given as in equation~\eqref{eq:flatstresstensor}, so that  under the~$T$-flow the Fourier coefficients~$\chi$ evolve as
\be
\frac{d\chi}{d\lambda} = - \frac{\pi^2}{96} k^3 D c_T \chi \, .
\ee
Thus for~$D > 0$, the~$\chi$ exponentially vanish and locally the flow converges to flat space, while for~$D < 0$ the~$\chi$ diverge, with smaller scales diverging faster, and the flow is locally ill-posed an in analogous manner to anti-diffusion.  
Hence the simplest dissipative dynamics on $\Mcal$ is actually ill-posed simply because~$\Sigma$ is unstable to ``crumpling'' on small scales. The well-posed~$T$-flow with $D > 0$ cannot obviously be viewed as a dynamics, but instead as a natural geometric flow for a $(2+1)$-dimensional CFT on a static curved space, which locally ``smoothes out'' the geometry.  Presumably, this flow will try to return the two-dimensional geometry $\Sigma$ to flat space or the round sphere critical points of $\Evac$ if the starting point is sufficiently near to these.

Thus in summary, the global properties of $\Evac$ appear to make dynamics driven by the vacuum energy pathological. Any dissipative dynamics will end in a singular geometry, and the most natural dynamics one might consider, namely local gradient descent of $\Evac$, is ill-posed on small scales leading to immediate ``crumpling'' of the geometry.

\section{Perturbative Bounds on Scalar Spectra}
\label{sec:spectrum}

Let us now extend away from bounds on~$\Evac$ and turn to the bound~\eqref{eq:specbound}, showing that the spectrum of any scalar operator in a holographic CFT on~$\rnum \times \Sigma$ is bounded below by the minimum Ricci scalar of~$\Sigma$.  Interestingly, in Section~\ref{subsec:holoscalars} we also showed that this same bound is obeyed by a free conformal scalar field.  It is natural to ask whether this bound is also universal: that is, does the spectrum of any conformal scalar operator obey~\eqref{eq:specbound}?  Here we will argue that -- at least in~$d = 3$ and when~$\Sigma$ is a perturbation of the round sphere -- it does.

First, consider a Euclidean CFT (for now in general~$d$) on~$\rnum \times \Sigma$ with~$\Sigma$ a perturbed~$S^{d-1}$, and assume this CFT contains a scalar operator~$\Ocal$.  We expect the two-point function of~$\Ocal$ to decay at large Euclidean time as
\be
\ev{\Ocal(\tau,\theta)\Ocal(\tau',\theta')} \sim e^{-\omega_\mathrm{min}|\tau-\tau'|},
\ee
where as before~$\theta$ and~$\theta'$ denote points on~$\Sigma$ and where for Lorentzian signature~$\omega_\mathrm{min}$ is the minimum frequency mode of~$\Ocal$ which we want to bound.  We may find such a bound by analyzing the behavior of this two-point function.  To that end, note that we may express this two-point function on~$\Sigma$ in terms of its value on the unperturbed~$S^{d-1}$ which, at leading order, is corrected via a three-point function with the stress tensor.  Writing the metric on~$\Sigma$ as~$\Omega_{ij} + \eps h_{ij}$ with~$\Omega_{ij}$ the metric on~$S^{d-1}$, we have
\begin{multline}
\ev{\Ocal(\tau,\Omega) \Ocal(\tau',\Omega')}_X = \ev{\Ocal(\tau,\theta) \Ocal(\tau',\theta')}_{S^{d-1}} \\
+ \frac{\epsilon}{2} \int d\tau'' \int_{S^{d-1}} d^{d-1}\theta'' \ev{\Ocal(\tau,\theta) \Ocal(\tau',\theta') T^{ij}(\tau'',\theta'')}_{S^{d-1}} h_{ij}(\theta'') 
+ O(\eps^2),
\end{multline}
where as usual the volume element on the~$S^{d-1}$ is understood.  Now, both the two-point function and three point-function above are universal, being determined only by the scaling dimension~$\Delta$ of~$\Ocal$~\cite{OsbPet93}. The two-point function is just the Weyl map of the flat space one, and corresponds to a spectrum of frequencies~$\omega^2$ that are algebraically determined by the~$SO(2,d)$ conformal symmetry.  Then considering the asymptotics of the leading correction in the large~$| \tau - \tau' |$ limit will confirm whether the bound~\eqref{eq:specbound} holds universally about the round sphere or not.

Now, since this leading contribution is universal we may conveniently compute it in a holographic setting.\footnote{For a direct, but more complicated CFT computation see the thesis of A. Hickling~\cite{HicklingThesis}. }  In fact, we may do better: rather than working with the two-point function and then extracting its large~$| \tau - \tau' |$ asymptotics, we may directly consider the leading~$O(\eps)$ correction to~$\omega_\mathrm{min}^2$ in the holographic case by considering the lowest fluctuation mode on the deformed bulk vacuum geometry dual to the perturbed boundary sphere.  This correction must reproduce the leading asymptotics computed from the above two-point function.  

Let us now restrict to~$d = 3$, as there the holographic computation is quite simple.   (In higher dimensions, general perturbations of the boundary metric contain independent tensor modes which then render the bulk solution more complicated.)  We perturb the boundary sphere as in~\eqref{eq:twospherepert} and again decompose the scalar perturbation~$\chi$ into spherical harmonics.  For a single mode~$Y_{\ell m}(\theta)$, the boundary metric takes the form
\be
ds^2_\Sigma = d\tau^2 + \left( 1 + \epsilon \, \mathrm{Re}\left(\chi_{\ell m} \, Y_{\ell m}(\theta) \right)\right) d\Omega_2^2
\ee
for a constant~$\chi_{\ell m}$; note that we have explicitly written out the real part to simplify notation in what follows.  Now, recall that for~$\ell=0$ and~$1$ this perturbation is simply a rigid scaling of the sphere or a rotation of it. Since we know the bound is saturated for a round sphere (and thus it will remain saturated for any rigid scaling or diffeomorphism), we neglect these cases and consider Appendix~\ref{app:flatpert} and~\ref{app:spherepert} only~$\ell>1$. Then from~\eqref{eq:sphereRicci} the right-hand side of the bound becomes
\begin{multline}
\label{eq:specboundrhs}
\frac{\Delta^2}{(d-1)(d-2)} R_\mathrm{min} = \\ \Delta^2 \left( 1 + \frac{\eps}{2} (\ell-1)(\ell+2) \min\mathrm{Re}\left(\chi_{\ell m} Y_{\ell m}\right) + O(\epsilon^2) \right) < \Delta^2 + O(\eps^2),
\end{multline}
where the inequality holds for any perturbation with~$\ell>1$ and we used the fact that~$\mathrm{Re}(\chi_{\ell m} Y_{\ell m})$ is oscillatory, so its minimum value is negative.

Next, let us obtain the perturbation to the left-hand side of the bound.  Following the discussion in Section~\ref{subsec:holoscalars} we consider a free massive scalar field~$\phi$ in the bulk dual to~$\Ocal$; then all we need to do is use~\eqref{eq:omegaexpr} to derive the perturbation to~$\omega_\mathrm{min}^2$.  In order to do so, we must rewrite the perturbed bulk metric~\eqref{eq:spherebulk} in the optical language of Section~\ref{sec:holog}, which can be done easily using the function~$Z = z/\sqrt{1+z^2}$ (which is unperturbed at~$O(\eps)$):
\begin{multline}
\overline{h}_{IJ}(y) dy^I dy^J =  \frac{ 1}{ \left( 1 + z^2 \right)^2} \left( dz^2 + \left( 1 + z^2 \right)\Omega_{ij} d\theta^i d\theta^j \right) \\
+ \eps \frac{1}{ \left( 1 + z^2 \right)} \left(A(z, \theta^i) \Omega_{ij} d\theta^i d\theta^j  +   B(z, \theta^i)    \frac{dz^2}{1+z^2} + 2  \partial_i C(z, \theta^i) d\theta^i dz \right) + O(\epsilon^2).
\end{multline}
It is then straightforward to show, using the solutions for~$A$,~$B$, and~$C$ found in Section~\ref{subsec:PTsphere}, that the optical Ricci scalar is
\be
\label{eq:Rbarpert}
\bar{R} = 6 \left( 1 + \eps \, \frac{b(z)}{z^2}  Y_{\ell m}(\theta^i) + O(\epsilon^2) \right),
\ee
with~$b(z)$ as given in~\eqref{eq:twospherefnb}.  Now, at leading order the lowest mode of the scalar has a constant wave function~$j_{\omega_\mathrm{min}}$, and thus from~\eqref{eq:omegaexpr}~$\omega^2_\mathrm{min} = \Delta^2$.  Including subleading corrections, this wave function must (up to an irrelevant overall constant) take the form
\be
\label{eq:jminpert}
j_{\omega_\mathrm{min}} = 1 + \eps \, \delta j + O(\eps^2);
\ee
then inserting~\eqref{eq:Rbarpert} and~\eqref{eq:jminpert} into~\eqref{eq:omegaexpr} (and bearing in mind that~$\ell > 1$ and that the integral of any~$Y_{\ell m}$ over the sphere vanishes for~$\ell \neq 0$), we find that~$\omega_\mathrm{min}^2$ is in fact unperturbed at~$O(\eps)$:
\be
\label{eq:specboundlhs}
\omega^2_\mathrm{min} = \Delta^2 + O(\epsilon^2).
\ee
Now comparing~\eqref{eq:specboundrhs} and~\eqref{eq:specboundlhs}, we find that (for~$d = 3$) the spectral bound~\eqref{eq:specbound} holds and \emph{is no longer saturated} at~$O(\eps)$.  We have shown this for any~$\ell>1$ harmonic perturbation to the unit round~$S^2$, and so clearly it remains true for any nontrivial sum over modes.  But due to the universality of the~$O(\eps)$ shift to~$\omega^2_\mathrm{min}$, this result must be true for \textit{any} CFT (and not just a holographic one).

Thus for any~$d=3$ CFT on~$\rnum \times \Sigma$ the bound~\eqref{eq:specboundrhs} must hold perturbatively when~$\Sigma$ is a perturbation of the round sphere.  Again we emphasize that for any nontrivial perturbation (so that the Ricci scalar $R$ is non-constant at order~$O(\eps)$), the bound is not saturated at order~$O(\eps)$: this implies that for sufficiently small perturbations of the sphere, higher order corrections in~$\eps$ (which will involve non-universal higher point functions) will not cause the bound to be violated; in other words, there must exist some neighborhood of the round sphere in which the bound holds.  The generality of this result (especially given that it holds nonperturbatively and in general dimension for both holographic field theories and a free conformal scalar) makes it quite natural to wonder whether this bound holds more generally than just in a neighborhood of the round sphere in~$d = 3$.

In particular, one might wonder how easily the argument provided here can be generalized to general dimension.  While some aspects of our argument are indeed valid for general~$d$, the key subtlety is that for~$d > 3$, perturbations of the boundary sphere have independent transverse-traceless tensor modes.  These (being non-scalar modes) cannot perturb either the boundary nor optical bulk Ricci scalars at order~$\eps$, and thus these tensor modes will leave the bound unaffected at order~$\eps$.  In particular, the bound will remain \textit{saturated} to~$O(\eps)$ for such perturbations of the round sphere, and therefore the behavior of higher-order corrections is crucial for determining whether or not the bound is obeyed.  But these higher-order corrections are not obviously fixed by conformal invariance.  For instance, the~$O(\eps^2)$ correction to~$\omega_\mathrm{min}^2$ will depend on the four-point function~$\ev{\Ocal\Ocal T T}$, which is constrained but not fixed by conformal invariance. Hence the universality arguments we exploited here would not apply.  Nevertheless, since only the asymptotics of the two-point function~$\ev{\Ocal\Ocal}$ are involved in the bound (rather than its full form), perhaps it is still possible to extract some universal behavior.  We leave this as an interesting direction for further study.

\section{Summary}
\label{sec:summary}

In this paper, we have explored whether certain properties of holographic CFTs on curved spaces are in fact universal to all CFTs.  We have focused on the negativity of the vacuum energy~$\Evac$ and a local generalization thereof for three-dimensional CFTs on~$\rnum \times \Sigma$ and have found that in any unitary CFT, if~$\Sigma$ is a static perturbation of flat space or a round sphere,~$\Evac$ (and its local generalization) must be negative.  A similar perturbative result holds for $\ln Z$ defined on deformations of a three-sphere, and this may be of relevance for holographic proposals for the universe wavefunction. 

By exploiting general features of~$\Evac$ we have extended this negativity result for~$\Evac$  away from the perturbative regime: the neighborhood of the round sphere or flat space on which~$\Evac$ is negative must in fact extend all the way to the boundary of the space of conformal geometries on which the CFT is well-defined.
This global extension allows us to conclude that any dynamical evolution of~$\Sigma$ governed solely by a minimization of~$\Evac$ must eventually become singular.  Indeed, the most natural dynamics, given by a gradient descent of $\Evac$, in fact is ill-posed in an analogous manner to anti-diffusion, where short distance behavior is pathological.  Interestingly the (unphysical) gradient ascent of~$\Evac$ gives a geometric flow of metrics on $\Sigma$ which acts to smooth inhomogeneity.

We have also investigated the universality of a bound on the spectrum of scalar CFT operators.  In particular, we have shown that in three dimensions and for perturbations of the round sphere, the minimum frequency~$\omega_\mathrm{min}^2$ of a scalar operator~$\Ocal$ is bounded from below by the minimum value of the Ricci scalar on~$\Sigma$.  Interestingly, this bound holds nonperturbatively and in any dimension not only in holographic CFTs with gravity duals, but also for the free conformal scalar.  Our perturbative results, coupled with the universality of the free \textit{and} strongly coupled regimes, are tantalizing hints that this result may indeed be broadly universal.

Perhaps the most interesting open questions are whether the negativity bounds on $\Evac$ and~$\rho_\mathrm{vac}$ apply universally to all three-dimensional CFTs for all metrics $\Sigma$ of flat or sphere topology.  For holographic CFTs this is true for both bounds (with certain caveats we have discussed, such as the existence of bulk metrics).  Moreover, we have argued that for any CFT the neighborhood of flat space or the round sphere about which~$\Evac$ is negative extends to the boundary of the space of metrics in some direction -- does it do so in all directions?  Likewise the scalar bound case is tantalizing as it is true non-perturbatively for a free conformal scalar CFT, and also holds in this and the holographic case in any dimension.  Does it also hold for all CFTs for all metrics $\Sigma$ of sphere topology?

The main theme in our analysis was to note that for perturbations of the round sphere or flat space, the bounds obtained from holographic CFTs can be rewritten in terms of two- and three-points functions of known field theoretic operators on flat space or on the sphere.  Because these two- and three-point functions are fixed (up to an overall constant) by conformal symmetry, they must take the same form in any CFT as they do in holographic ones, and thus the holographic bounds must hold (perturbatively) in all CFTs.  While this argument is quite simple from the field theoretic perspective, a priori this universality was not at all obvious.  It was only thanks to the universal \textit{holographic} results outlined in Section~\ref{sec:holog} that we we able to obtain results universal to all CFTs.  Thus  the work presented here is an example of holographic intuition leading to novel and universal field theoretic findings\footnote{Indeed, another universal holographic bound is that of~\cite{Fischetti:2016fbh} regarding entanglement entropy; it would be interesting if this could be shown to be universal to all CFTs as well.}.  It is quite possible (indeed, likely) that additional universal CFT results lie waiting to be discovered in this way; we leave these discoveries for exciting future work.

%=============================== Acknowledgements  ==================================
\section*{Acknowledgements}
\label{sec:thanks}

We would like to thank both Andrew Hickling and Paul McFadden for initial collaboration and many useful discussions. We also thank Alexandre Belin, Henry Maxfield, and Slava Rychkov for valuable discussions as well as Sergi Solodukhin whom we thank for the term ``$T$-flow'', and Juan Maldacena who pointed out to TW the relevance of the Gauss-Bonnet bound on renormalized bulk volume.  This work was supported by STFC grant ST/L00044X/1. SF also thanks the University of California, Santa Barbara for hospitality while some of this work was completed.  

\appendix

\section{Deforming Euclidean Flat Space}
\label{app:flatpert}

In this Appendix, we extend the computations of Section~\ref{sec:PT} to consider the behavior of the partition function under perturbations of flat Euclidean space.  We will perform the computation both in field theory and holographic language in order to show that the results agree (as they must).

\subsection{CFT Calculation}
\label{subapp:CFT}

Consider the flat Euclidean metric~$G_{AB} = \delta_{AB}$.  A general perturbation can be decomposed into scalar, vector, and tensor parts with respect to the flat background as
\be
\label{eq:SVTflat}
H_{AB} = \chi \, \delta_{AB} + \partial_{(A} v_{B)} + H^{TT}_{AB},
\ee
where~$H^{TT}_{AB}$ is transverse and traceless. The scalar part of this perturbation simply corresponds to a perturbative Weyl scaling of the metric, and hence must leave the partition function invariant.  Likewise, the vector part just generates infinitesimal diffeomorphisms, and must also leave the partition function unchanged.  Hence the only physical perturbations are the TT tensor modes.  We then decompose the perturbation into Fourier modes as
\be
\label{eq:EucTTpert}
H^{TT}_{AB}(x) = \int d^3k \, a_{AB}(k) e^{-i k \cdot x}
\ee
where the transverse-traceless conditions imply that the Fourier coefficients~$a_{AB}$ obey $a_{AB} \delta^{AB} = 0$,~$k^A a_{AB} = 0$, while reality of the perturbation requires~$a_{AB}(-k) = a_{AB}^*(k)$.  Likewise, in Fourier space the universal form~\eqref{eq:TTuniversal} of the two-point function of the stress tensor becomes~\cite{OsbPet93,Bzowski:2013sza},
\begin{subequations}
\be
\label{eq:TTuniversalFourier}
\ev{T_{AB}(x) T_{CD}(y)}_0 = \int d^3k \, d^3p f_{AB,CD}(k,p)  e^{- i ( k \cdot x + p \cdot y)},
\ee
where
\bea
f_{AB,CD}(k,p) &\equiv \frac{1}{(2\pi)^3} \frac{c_T \pi^2}{24} k^3 \left(\pi_{A(C} \pi_{D)B} - \frac{1}{2} \pi_{AB} \pi_{CD} \right) \delta( k + p ), \\
\pi_{AB} &\equiv \delta_{AB} - \frac{k_A k_B}{k^2}.
\eea
\end{subequations}
Note that to regulate the UV divergence of the two-point function, we have used
\be
\int d^3x  \frac{1}{x^p}  \, e^{- i k \cdot x} = 4 \pi k^{p-3} \Gamma\left[ 2 - p \right] \sin\left( \frac{ p \pi}{2} \right),
\ee
which converges for~$1 < p < 3$, and analytically continued it to~$p=6$ to obtain
\be
\int d^3x  \frac{1}{x^6}  \, e^{- i k \cdot x} = \frac{\pi^2}{12} k^{3}.
\ee
Then inserting the UV-regulated~\eqref{eq:TTuniversalFourier} into~\eqref{eq:lnZvariation} we obtain
\be
\label{eq:flatvariation}
\ln Z[\delta + \eps H] = \ln Z[\delta]  + \eps^2 \frac{c_T \pi^5}{24} \int d^3k \, k^3\, \left| a_{AB}(k) \right|^2 + O(\eps^3).
\ee
Thus we find that for any physical perturbation (i.e.~excluding the constant~$k = 0$ mode which just corresponds to a coordinate rescaling), the partition function \textit{increases} from its unperturbed value at order~$\eps^2$:
\be
\delta_\eps^2 \ln Z[\delta + \eps H] > 0.
\ee

\subsection{Holographic Calculation}
\label{subapp:holo}

As a check of this universality, we may obtain this same result from a holographic calculation.  In this case, we will compute the one-point function of the stress tensor, which we may then insert into~\eqref{eq:linvariation2}.

The holographic dual to flat Euclidean space is hyperbolic space, so we consider a perturbation to the bulk of the form
\be
ds^2_\mathrm{bulk} = \frac{l^2}{z^2} \left[ dz^2 + \delta_{AB} dx^A dx^B  + \eps F_{AB}(z,x) dx^A dx^B \right],
\ee
where we have used the local coordinate freedom to chose a Feffermann-Graham gauge.  Clearly the boundary perturbation is~$H_{AB}(x) = F_{AB}(0,x)$, so for the physical perturbations in which we are interested,~$F_{AB}$ must be transverse and traceless at the boundary.  The linearized bulk constraint equations then yield
\be
\partial_z \xi_A = 0, \quad \frac{1}{2} \partial^A \xi_A + \frac{1}{z} \partial_z F = 0,
\ee
where~$\xi_A \equiv \partial^B F_{AB} - \partial_A F$ and~$F \equiv \delta^{AB} F_{AB}$.  Since~$F_{AB}$ is transverse and traceless at the boundary, the constraint equations above in fact imply that~$F_{AB}$ must be transverse and traceless everywhere in the bulk.  The linearized bulk dynamical equation then becomes
\be
\label{eq:flatEuc}
\partial^2 F_{AB} + \partial_z^2 F_{AB} - \frac{2}{z} \partial_z F_{AB}  = 0,
\ee
whose regular solution at~$z \to \infty$ is
\be
F_{AB}(z,x) = \int d^3k \, a_{AB}(k) e^{-i k \cdot x} (1 + k z) e^{-k z}
\ee
Using the standard prescription~\cite{BalKra99,deHSol00} we may obtain the boundary stress tensor as the cubic term in an expansion of~$F_{AB}$ about~$z = 0$.  We then obtain
\be
\ev{T_{AB}}_{\delta + \eps H} = \eps \frac{\pi^5 c_T}{48} \int d^3k \, k^3 a_{AB}(k) e^{- i k \cdot x} + O(\eps^2),
\ee
where we used the relation~\eqref{eq:cTgrav}.  Plugging this expression back into~\eqref{eq:linvariation2} we recover the universal variation~\eqref{eq:flatvariation}, as we must.

\section{Deforming the Three-Sphere}
\label{app:spherepert}

In this Appendix, we extend the computations of Section~\ref{sec:PT} to consider the behavior of the partition function under perturbations of the round three-sphere~$S^3$.  In principle, we could obtain this behavior by an appropriate Weyl rescaling of the results in Appendix~\ref{app:flatpert}, since the three-sphere is locally Weyl flat.  In fact, it follows immediately from the results of Appendix~\ref{app:flatpert} that the partition function must increase under perturbations of the sphere that vanish (sufficiently rapidly) at the pole.  To see this, it is sufficient to note that the three-sphere metric~$\Omega_{AB}$ can be obtained from a Weyl rescaling of flat space which is singular at the pole:~$\Omega_{AB} = \omega^2(x) \delta_{AB}$, with
\be
\omega(x) = \frac{2}{1+|x|}
\ee
in standard Cartesian coordinates, and the point~$|x| \to \infty$ at which~$\omega$ vanishes corresponds to the north pole of the sphere.  Then any perturbation of flat space~$H_{AB}$ can be mapped to a perturbation of the sphere by the same Weyl rescaling:~$H^\mathrm{(sphere)}_{AB} = \omega^2(x) H_{AB}$; note that if~$H_{AB}$ is regular on flat space, then~$H^\mathrm{(sphere)}$ must vanish at the north pole of the sphere.  However, since the geometries~$\delta_{AB} + \eps H_{AB}$ and~$\Omega_{AB} + \eps H^\mathrm{(sphere)}_{AB}$ are related by a Weyl rescaling, their partition functions must agree.  Thus since we found that~$\delta_\eps^2 \ln Z[\delta + \eps H] > 0$, we must also have that~$\delta_\eps^2 \ln Z[\Omega + \eps H^\mathrm{(sphere)}] > 0$.  In other words, the partition function increases (at order~$\eps^2$) under perturbations of the three-sphere which arise from a Weyl rescaling of regular perturbations of Euclidean space.

What about more general perturbations of the sphere?  In principle one could obtain such perturbations by an appropriate limit of regular perturbations of flat space, and therefore the same result would apply.  However, we may circumvent this issue entirely by performing the calculation of~$\delta_\eps^2 Z$ in the holographic setting, after which the arguments of Section~\ref{sec:PT} ensure the result must be universal.

We thus proceed similarly to the flat space computation.  First we decompose a general metric perturbation on the three-sphere as
\be
\label{eq:SVTsph}
H_{AB}(x) = \chi \, \Omega_{AB} + \grad_{(A} v_{B)} + H^{TT}_{AB}(x),
\ee
where again~$H^{TT}_{ij}$ is transverse and traceless.  As in flat space, the scalar and vector perturbations are physically uninteresting and so we ignore them.  We then decompose~$H^{TT}_{AB}$ into transverse-traceless tensor spherical harmonics~\cite{San78}:
\be
\label{eq:tensorsphereexpansion}
H^{TT}_{AB}(\theta^C) = \sum_{n \ell m} a_{n \ell m} Y^{n \ell m}_{AB}(\theta^C),
\ee
where reality of~$H^{TT}_{AB}$ implies~$a_{n \ell,-m} = a_{n \ell m}^*$.  For reference, we remind the reader that for transverse-traceless tensor harmonics,~$n \geq 1$ and
\be
\grad^2 Y^{n \ell m}_{AB} = -(n(n+2)-2) Y^{n \ell m}_{AB}.
\ee

We now look for a bulk solution: we write the bulk metric as
\be
ds^2_\mathrm{bulk} = \frac{l^2}{z^2} \left[\frac{dz^2}{1+ z^2} + \Omega_{AB} d\theta^A d\theta^B  + \eps F_{AB}(z,\theta) d\theta^A d\theta^B \right]
\ee
from which it follows that~$H^{TT}_{AB} = F_{AB}(0,\theta)$.  The linearized constraint equations yield
\be
\partial_z \xi_A = 0, \quad \frac{1}{2} \grad^A \xi_A + \frac{1 + z^2}{z} \partial_z F - F = 0,
\ee
where~$\xi_A \equiv \grad^B F_{AB} - \partial_A F$ and~$F \equiv \Omega^{AB} F_{AB}$ and indices are raised and lowered by the round sphere metric~$\Omega_{AB}$.  Again we find that the constraint equations require~$F_{AB}$ to be transverse-traceless everywhere in the bulk (since it is at the boundary).  Then for a single tensor spherical harmonic mode~$F_{AB} = f_{n\ell m}(z) Y^{n\ell m}_{AB}$, the linearized dynamical bulk equation yields
\be
\left( 1 + z^2 \right) \partial_z^2 f_{n\ell m} - \frac{2 + z^2}{z} \partial_z f_{n\ell m} - n(n+2) f_{n\ell m} = 0,
\ee
whose regular solution is
\begin{multline}
f_{n\ell m}(z) = {}_2F_1\left[ -\frac{n+2}{2}, \frac{n}{2}, -\frac{1}{2} , -z^2 \right] \\
 + \frac{n(n+1)(n+2)}{3} \, {}_2F_1\left[\frac{1-n}{2} , \frac{n+3}{2}, \frac{5}{2} , -z^2 \right] z^3.
\end{multline}

Note that~$f_{n \ell m}$ has the near-boundary expansion
\be
f_{n \ell m} = 1 - \frac{n(n+2)}{2} \, z^2 + \frac{n(n+1)(n+2)}{3} \, z^3 + \cdots.
\ee
The stress tensor is obtained in the usual way from the third-order coefficient of this expansion\footnote{In principle we should first convert to a Fefferman-Graham coordinate~$Z$, defined by
\be
Z(z) = \frac{z}{1+\sqrt{1+z^2}},
\ee
but this does not affect the coefficient of the relevant term in the expansion of~$f_{n \ell m}$.}; thus for the metric perturbation~\eqref{eq:tensorsphereexpansion}, we obtain
\be
\ev{T_{AB}}_{\Omega + \eps H} = \eps \frac{\pi^2 c_T}{48} \sum_{n \ell m} n(n+1)(n+2) a_{n \ell m} Y^{n \ell m}_{AB} + O(\eps^2).
\ee
Using the orthonormality of tensor spherical harmonics,
\be
\int_{S^3} \Omega^{AC} \Omega^{BD} Y_{AB}^{n\ell m} {Y_{CD}^{n'\ell'm'}}^* = \int_{S^3} \Omega^{AC} \Omega^{BD} Y_{AB}^{n\ell m} Y_{CD}^{n'\ell',-m'} = \delta_{nn'} \delta_{\ell\ell'} \delta_{mm'},
\ee
we may plug our stress tensor and metric perturbation into~\eqref{eq:linvariation2} to obtain
\be
\ln Z[\Omega + \eps H] = \ln Z[\Omega] + \frac{\eps^2}{4} \frac{\pi^2 c_T}{48} \sum_{n\ell m} n(n+1)(n+2) \left|a_{n\ell m}\right|^2 + O(\eps^3).
\ee
Thus as desired (and recalling that~$n \geq 1$ for transverse-traceless tensor spherical harmonics), we see that any perturbation of the three-sphere increases~$\ln Z$ from its value on the round sphere at leading nontrivial order in~$\eps$.

\bibliographystyle{jhep}
\bibliography{all}

\end{document}